\documentclass[journal,lettersize]{IEEEtran}

\usepackage{aligned-overset}
\usepackage{hyperref}
\usepackage{graphicx}
\usepackage{makecell}
\usepackage{multirow}
\usepackage{tabularx}
\usepackage{multirow}
\usepackage{tabularx}
\usepackage{enumerate}
\usepackage{bbm}
\usepackage{latexsym}
\usepackage{wasysym}
\usepackage{subfigure}
\usepackage{amssymb}
\usepackage{comment}

\newcommand\noarxiv[1]{#1}

\newtheorem{thm}{\textbf{\textsc{Theorem}}}

\begin{document}

\title{UPPRESSO: Untraceable and Unlinkable Privacy-PREserving Single Sign-On Services}

\author{Chengqian Guo, Jingqiang Lin, Quanwei Cai, Wentian Zhu, Wei Wang, Jiwu Jing, Qiongxiao Wang, Bin Zhao, and Fengjun Li
\thanks{Chengqian Guo is affiliated with Yuncheng Vocational and Technical University, China. Jingqiang Lin (corresponding author, linjq@ustc.edu.cn), Wentian Zhu, and Wei Wang are affiliated with University of Science and Technology of China (USTC), China. Quanwei Cai is with Beijing Zitiao Network Technology Co., Ltd, China. Jiwu Jing is with University of Chinese Academy of Sciences (UCAS), China. Qiongxiao Wang is with Beijing Certification Authority (BJCA) Co., Ltd, China. Bin Zhao is with Tencent America, USA. Fengjun Li is with The University of Kansas, USA.}
}



\maketitle



\begin{abstract}
Single sign-on (SSO) protocols such as OpenID Connect (OIDC),
allow a user to maintain only the credential for an identity provider (IdP) to log into multiple relying parties (RPs).
However, SSO introduces privacy threats, as ({\em a}) a curious IdP could trace a user's visits to RPs, and ({\em b}) colluding RPs could learn a user's online profile by linking her identities across these RPs.
This paper presents a privacy-preserving SSO scheme, called UPPRESSO, to protect a user's online profile against both (\emph{a}) an honest-but-curious IdP and (\emph{b}) malicious RPs colluding with some users.
An identity-transformation approach is proposed to generate untraceable ephemeral pseudo-identities for an RP and a user in each login, 
 from which the visited RP derives a permanent account for the user, 
  while these identity transformations provide unlinkability among the logins visiting different RPs.
We built the UPPRESSO prototype providing OIDC-compatible SSO services, accessed from a commercial-off-the-shelf browser without installing any extension.
The extensive evaluations show that it fulfills the security and privacy requirements of SSO with reasonable overheads.
\end{abstract}

\maketitle


\section{Introduction}
\label{sec:intro}
Single sign-on (SSO) protocols such as OpenID Connect (OIDC) \cite{OpenIDConnect}, OAuth 2.0 \cite{rfc6749}, and SAML \cite{SAML, SAMLIdentifier}, are widely deployed for identity management and authentication.
SSO allows a user to log into a website from browsers,
 known as the \emph{relying party} (RP), using her account registered at another trusted web service, known as the \emph{identity provider} (IdP).
An RP delegates its user identification and authentication to the IdP, which issues an \emph{identity token} (such as ``id token'' in OIDC and ``identity assertion'' in SAML) for a user to visit the RP.
For instance, when an OIDC user initiates a login to visit an RP,
the target RP constructs an identity-token request with its identity and redirects it to a trusted IdP.
After authenticating this user, the IdP issues an identity token binding the identities of the authenticated user and the target RP (i.e., $ID_U$ and $ID_{RP}$), which is returned to the user and then forwarded to the RP.
Finally, the RP verifies the identity token to determine whether the token holder is allowed to log in. Thus, a user maintains only one credential for user authentication to the IdP, instead of multiple credentials for visiting different RPs.


The wide adoption of SSO raises concerns on user privacy, because it facilitates the tracking of a user's login activities  \cite{NIST2017draft, SPRESSO, BrowserID, maler2008venn}. 
To issue identity tokens, an IdP should know the target RP to be visited by a user.
So a curious IdP could trace a user's all login activities
 \cite{BrowserID, SPRESSO},
called {\em IdP-based login tracing} in this paper.
Another privacy threat arises from the fact that RPs learn a user's identity from the identity tokens they receive.
If an \emph{identical} user identity is enclosed in the tokens for a user to visit different RPs, colluding RPs could link the logins across these RPs 
to learn the user's online profile \cite{maler2008venn, FirefoxAccount}.
This is called {\em RP-based identity linkage}. 

This paper presents UPPRESSO, an Untraceable and Unlinkable Privacy-PREserving Single Sign-On protocol.
We propose {\em identity transformations} for SSO services,
 and integrate them into OIDC.
When visiting an RP in UPPRESSO, a user transforms $ID_{RP}$ into \emph{ephemeral} $PID_{RP}$, which is sent to an IdP to transform $ID_U$ into an \emph{ephemeral} user pseudo-identity $PID_U$.
Then, the identity token issued by the IdP binds only $PID_U$ and $PID_{RP}$, instead of $ID_U$ and $ID_{RP}$. On receiving the token, 
 the target RP transforms $PID_U$ into an account \emph{unique at each RP} but \emph{identical across multiple logins} visiting this RP.

\begin{figure*}[tb]
  \centering
  \includegraphics[width=0.817\linewidth]{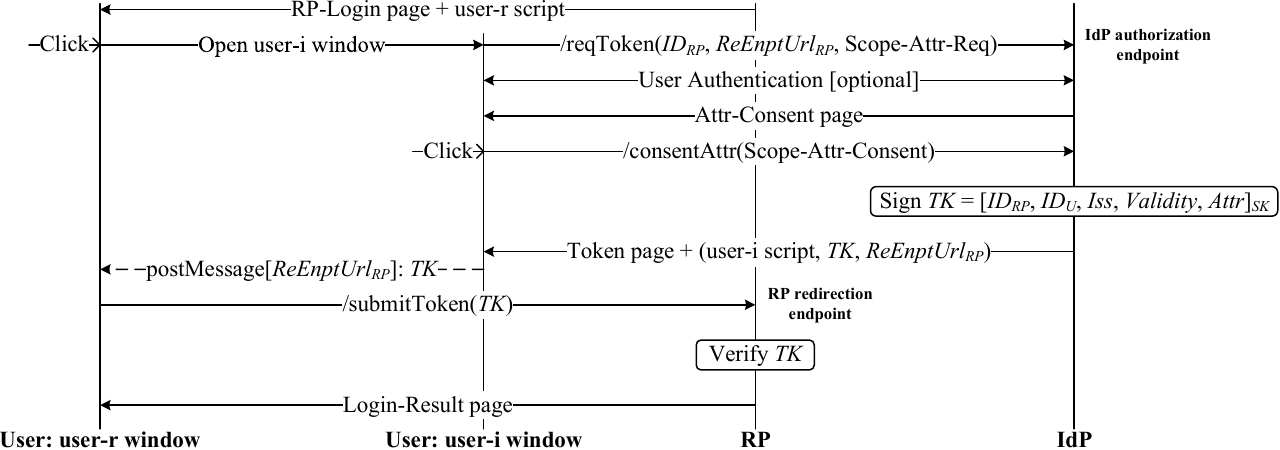}
  \label{fig:OpenID}
  \caption{The implicit SSO login flow of OIDC with pop-up UX}
\end{figure*}

UPPRESSO prevents both IdP-based login tracing and RP-based identity linkage, while existing privacy-preserving SSO schemes address only one of them \cite{BrowserID, SPRESSO, NIST2017draft, save-flow,POIDC} or introduce another fully-trusted component in addition to the IdP \cite{miso,SPRESSO,up-sso}.
In Section \ref{sec:background} we analyze these solutions,
    and also compare UPPRESSO with privacy-preserving identity federation \cite{PseudoID, ELPASSO, UnlimitID, Opaak, uprov, hyperledge-idemix} where browser extensions are required.
UPPRESSO offers an easy-to-deploy option to enhance the protections of user privacy for SSO, because no extra component is introduced and the services are accessed from commercial-off-the-shelf (COTS) browsers without extensions.
Our contributions are as follows.
\begin{itemize}
\setlength{\topsep}{0pt}
\setlength{\partopsep}{0pt}
\setlength{\itemsep}{0pt}
\setlength{\parsep}{0pt}
\setlength{\parskip}{0pt}
\item We proposed a novel identity-transformation approach for privacy-preserving SSO services and designed identity-transformation algorithms with desirable properties.
\item We designed the UPPRESSO protocol by integrating the identity transformations into OIDC compatibly, and proved the security and privacy guarantees.
\item We implemented the UPPRESSO prototype on top of an open-source OIDC implementation. Through performance evaluations, we confirmed that UPPRESSO introduces reasonable overheads.
\end{itemize}

We explain the privacy threats in SSO and present related works in Section \ref{sec:background}.
The identity-transformation approach is proposed in Section \ref{sec:challenge}, followed by the detailed designs in Section \ref{sec:UPPRESSO}.
Security and privacy are analyzed in Section \ref{sec:analysis}.
We present the prototype implementation and evaluations in Section \ref{sec:implementation}, and discuss extended issues in Section \ref{sec:discussion}. Section \ref{sec:conclusion} concludes this work.

\section{Background and Related Work}
\label{sec:background}

\subsection{OpenID Connect}
\label{subsec:OIDC}
OIDC \cite{OpenIDConnect} is the most popular SSO protocol. It supports different login flows: implicit flow, authorization code flow, and hybrid flow (a mix of the other two).
The flows differ in the steps for requesting, receiving and forwarding identity tokens, but have common security requirements for the tokens.
We present our designs in the implicit flow and discuss the support for the authorization code flow in Section \ref{sec:discussion}.

Users and RPs register at an IdP with their identities
and other information such as user credentials (e.g., passwords)
and RP \emph{redirection endpoint URLs} to receive tokens (i.e., $ReEnptUrl_{RP}$), respectively  \cite{rfc6749}.
denoted as $ReEnptUrl_{RP}$, respectively  \cite{rfc6749}.

User operations in the login flows are conducted on a user agent (or browser typically).
The flow with pop-up UX is proposed to provide an in-context user experience \cite{dimvaLiM16,GoogleIdIntegrate,uber}.
As shown in Figure \ref{fig:OpenID},
when a user clicks the SSO login button in a browser window visiting the target RP 
(called the \emph{user-r window}),
 an identity-token request is constructed with the RP's identity, the RP's redirection endpoint URL, and the scope of requested user attributes.
It pops up another window (i.e., the \emph{user-i window}),
    and this request is sent to the IdP's \emph{authorization endpoint} where tokens are issued \cite{rfc6749,OpenIDConnect}.
After authenticating the user, the IdP issues an identity token that encloses the (pseudo-)identities of the authenticated user and the target RP, the user attributes, a validity period, etc.
The token is then forwarded to the RP's redirection endpoint via the user-i window.
Finally, the RP verifies the token and allows the token holder to log in as the enclosed user (pseudo-)identity.
There are scripts in two browser windows (called the \emph{user-r script} and the \emph{user-i script}),
    responsible for communicating with the respective origin web servers
    and also the cross-origin
communications by the \verb+postMessage+ HTML5 API \cite{postm-targeto} within the browser.


Alternatively, if a browser works as the user agent of OIDC but with only one window,
        user attributes are consented in this window when redirected to the IdP \cite{OpenIDConnect,rfc6749,GoogleIdIntegrate}.
        This is called \emph{redirect UX}.
After a token is issued by the IdP,
    it is sent to the RP's redirection endpoint as a fragment (but not query parameters) of the URL due to security considerations,
so a script from the RP is needed to handle this \cite{de2014oauth}.
The UPPRESSO prototype is built with pop-up UX,
    and we discuss the support for redirect UX in Section \ref{sec:discussion}.

\begin{table*}[tb]
    \caption{Privacy-preserving solutions of SSO and identity federation}
    \centering{
    \begin{tabular}{|c|c|c|c|c|c|}
  \hline
  & \multirow{2}*{\textbf{Solution}} &
  \multicolumn{2}{c|}{\textbf{Privacy Threat}$^{\sharp}$} & \textbf{Extra Trusted} & \textbf{Accounts Derived from} \\ \cline{3-4}
  & & IdP-based Login Tracing & RP-based Identity Linkage & \textbf{Component}$^{\dag}$ & \textbf{User Secret}$^{\ddag}$ \\\hline

  \multirow{6}*{SSO} & OIDC w/ PPID \cite{NIST2017draft} & $\Circle$ & $\CIRCLE$ & $\CIRCLE$ & $\CIRCLE$ \\ \cline{2-6}
   & BrowserID \cite{BrowserID} & $\CIRCLE$ & $\Circle$ & $\CIRCLE$ & $\CIRCLE$ \\ \cline{2-6}
   & SPRESSO \cite{SPRESSO} & $\CIRCLE$ & $\Circle$ & $\LEFTcircle$$^1$ & $\CIRCLE$ \\ \cline{2-6}
   & POIDC \cite{POIDC,save-flow} & $\CIRCLE$ & $\Circle$$^2$ & $\CIRCLE$ & $\CIRCLE$ \\ \cline{2-6}
   & MISO \cite{miso} & $\CIRCLE$ & $\CIRCLE$ & $\Circle$ & $\CIRCLE$ \\ \cline{2-6}
   & UP-SSO \cite{up-sso} & $\CIRCLE$ & $\CIRCLE$ & $\CIRCLE$ & $\Circle$$^3$ \\ \hline 
  \multirow{5}*{\makecell{Identity\\Federation}} & PRIMA \cite{prima} & $\CIRCLE$ & $\Circle$ & $\CIRCLE$ & $\Circle$ \\ \cline{2-6}
   & PseudoID \cite{PseudoID} & $\CIRCLE$ & $\CIRCLE$ & $\Circle$ & $\Circle$ \\ \cline{2-6}
   & Opaak \cite{Opaak} & $\CIRCLE$ & $\CIRCLE$ & $\CIRCLE$ & $\Circle$ \\ \cline{2-6}
   & PP-IDF \cite{ELPASSO,uprov,UnlimitID} & $\CIRCLE$ & $\CIRCLE$ & $\CIRCLE$ & $\Circle$ \\ \cline{2-6}
   & Fabric Idemix \cite{hyperledge-idemix} & $\CIRCLE$ & $\CIRCLE$ & $\CIRCLE$ & $\Circle$ \\ \hline
  SSO & UPPRESSO & $\CIRCLE$ & $\CIRCLE$ & $\CIRCLE$ & $\CIRCLE$ \\ \hline
\end{tabular}}
    \label{tbl:comparison-protocol}
{
\begin{itemize}
\item[]
${\sharp}$ \textbf{Privacy Threat}: $\CIRCLE$ prevented, $\Circle$ not prevented.
\ \ \ \ \ \ \ \ \ \ \ \ \ \ \ \ \ \ \ \ \ \ \ \ \ \ \ \ \ \ ${\dag}$ \textbf{Extra Trusted Component}: $\CIRCLE$ no, $\Circle$ required.\\
${\ddag}$ \textbf{Accounts Derived from User Secret}: $\CIRCLE$ no, $\Circle$ yes.

\item[1.]
SPRESSO assumes a malicious IdP, so a trusted forwarder server is introduced to decrypt the RP identity and forward tokens to the RP.

\item[2.]
A variation of POIDC \cite{POIDC} proposes to also hide a user's identity in the commitment and prove this to the IdP in zero-knowledge,
        but it takes seconds or more to generate such a zero-knowledge proof (ZKP) even for a powerful server \cite{ZKP-BINF,zkp-benchmark,ZKP-GPU},
        which is impracticable for a user agent in SSO.
\item[3.]
UP-SSO \cite{up-sso} runs a user-side Intel SGX enclave with a secret to calculate PPIDs.
%
\end{itemize}}
\end{table*}

\subsection{Privacy-Preserving SSO and Identity Federation}
\label{subsec-solutions}
In the original OIDC protocol 
    an IdP signs identity tokens binding the identities of the authenticated user and the target RP \cite{OpenIDConnect,rfc6749}.
A user's login activities are leaked due to these identities:
        The curious IdP learns the visited RP from $ID_{RP}$ enclosed in an identity-token request (i.e., IdP-based login tracing),
        while colluding RPs link a user's logins visiting these RPs by $ID_{U}$ in the tokens (i.e., RP-based identity linkage).


\noindent\textbf{Privacy-preserving SSO.}
Existing solutions propose various mechanisms to hide (\emph{a}) RP identities from the IdP and (\emph{b}) user identities from the colluding RPs.
Table \ref{tbl:comparison-protocol} compares these schemes,
and only UPPRESSO provides services accessed from COTS browsers while preventing both of the two privacy threats and not introducing extra trusted components in addition to the honest IdP.

Pairwise pseudonymous identifiers (PPIDs) are specified  to protect user privacy against colluding RPs \cite{OpenIDConnect, SAMLIdentifier,NIST2017draft}.
An IdP assigns a unique PPID for a user to log into every RP and encloses it in identity tokens, so colluding RPs cannot link the user by PPIDs.
It does not prevent IdP-based login tracing, because the IdP needs the RP's identity to assign PPIDs.

Other privacy-preserving schemes prevent IdP-based login tracing but leave users vulnerable to RP-based identity linkage, due to the identical user identities enclosed in identity tokens.
For example, in BrowserID \cite{BrowserID} 
after authenticating a user,
    an IdP 
issues a ``user certificate'' binding the user's identity to an ephemeral public key.
The user then uses the corresponding private key to sign a subsidiary ``identity assertion'' that binds the target RP's identity and sends both of them to the RP.
In SPRESSO an RP creates a one-time tag (i.e., ephemeral pseudo-identity) for each login \cite{SPRESSO},
 and in POIDC \cite{POIDC,save-flow} a user sends an identity-token request with a commitment on the target RP's identity (but not with plain $ID_{RP}$),
        which are enclosed in identity tokens along with $ID_U$.


MISO \cite{miso} decouples the assignment of PPIDs from an honest IdP,
 to an extra fully-trusted mixer server that calculates a user's PPID based on $ID_U$, $ID_{RP}$ and a secret after it receives the user's identity from the IdP.
MISO prevents RP-based identity linkage for the RPs receives only PPIDs,
    and IdP-based login tracing for $ID_{RP}$ is disclosed to the mixer but not the IdP.
It protects user privacy against even collusive attacks by the IdP and RPs,
    but the mixer could track a user's all login activities.
Similarly, UP-SSO \cite{up-sso} runs a fully-trusted Intel SGX enclave on the user side,
 which is remotely attested by the IdP and receives the secret to generate PPIDs. 

\noindent\textbf{Privacy-preserving identity federation.}
Identity federation enables a user registered at a trusted IdP to be accepted by other parties, potentially with different accounts,
but 
a long-term user secret is involved in the derivation of accounts,\footnote{Although the term ``single sign-on (SSO)'' was used in some schemes \cite{PseudoID, Opaak, ELPASSO, WangWS13, HanCSTW18, HanCSTWW20}, this paper refers to them as \emph{identity federation} to emphasize the difference.
Such a user-maintained secret used to derive accounts in identity federation harm the user convenience:
Once it is lost or leaked, the user has to notify all RPs to update her accounts derived from this secret 
    and then the ``singleness'' property is broken,
while SSO aims to minimize a user's interactions with RPs.
Note that this user secret involved in the derivation of accounts is very different from the credential for user authentication to the IdP: If this credential is lost or leaked in SSO, a user  renews it at the IdP only.}
 so browser extensions are required to maintain this user secret.


In PRIMA \cite{prima}, an IdP signs a credential
that binds user attributes and  a verification key.
Using the corresponding signing key, the user authenticates herself and provides selected attributes to RPs. The verification key works as the user's global identity and exposes her to RP-based identity linkage.
PseudoID \cite{PseudoID} introduces another trusted server in addition to the trusted IdP,
 to blindly sign \cite{blind-sign}
a token that binds a long-term user secret and a pseudonym.
The user then unblinds this token,
        presents it to the IdP which processes the secret,
    and then authenticates herself to an RP by this secret.

Privacy-preserving identity federation schemes are designed based on anonymous credentials \cite{anon-credential-2001, idemix, anon-credential}.
In Opaak \cite{Opaak}, UnlimitID \cite{UnlimitID}, U-Prove \cite{uprov}, and EL PASSO \cite{ELPASSO}, the IdP signs anonymous credentials, which bind a user secret, 
and a user proves ownership of the anonymous credentials using her secret.
Similarly, Fabric \cite{hyperledge-idemix} integrates Idemix anonymous credentials \cite{idemix} for unlinkable pseudonyms. 

Some privacy-preserving solutions of identity federation \cite{PseudoID,ELPASSO,UnlimitID,Opaak,uprov,hyperledge-idemix} prevent both IdP-based login tracing and RP-based identity linkage, for (\emph{a}) the RP's identity is not enclosed in the anonymous credentials and (\emph{b}) the user selects different pseudonyms to visit different RPs.
They even protect user privacy against collusive attacks by the IdP and RPs, as these pseudonyms cannot be linked through anonymous credentials \cite{anon-credential-2001, idemix, anon-credential} when the ownership of these credentials is proved to colluding RPs. 

%

\noindent\textbf{Anonymous identity federation.}
Such approaches offer the strongest privacy,
    as a user visits RPs with pseudonyms that cannot be exploited to link any two actions.
Anonymous identity federation was formalized and constructed using cryptographic primitives (e.g., ZKP and group signature) \cite{HanCSTWW20, HanCSTW18, WangWS13}.
Extended features including proxy re-verification \cite{HanCSTWW20}, designated verification \cite{HanCSTW18} and distributed IdP servers \cite{TSAPP}, are also considered. 
These completely-anonymous authentication services only work for special scenarios and do not support user identification (or account uniqueness) at each RP, a common requirement in most applications.

\subsection{OPRF-based Applications}
PrivacyPass \cite{privacypass} and TrustToken \cite{trusttoken} adopted the oblivious pseudo-random function (OPRF) of Hash(EC)DH \cite{oprf-proved,voprf-proved} to sign anonymous tokens. 
A user generates a random number $e$ to blind an unsigned token $T$ into $[e]T$. After authenticating the user, a token server signs $[e]T$ using its private key $k$ and returns $[k e]T$. The user then uses $e$ to convert it into ($T, [k]T$),
        which is redeemed when she anonymously accesses resources.
In this procedure, the token server obliviously calculates a pseudo-random output as an anonymous token,
    because ($T, [k]T$) and ($[e]T, [ke]T$) cannot be linked.


UPPRESSO employs a similar cryptographic technique (see Section \ref{subsec:overview} for details).
 A user selects $t$ to transform $ID_{RP}$ into $PID_{RP} = [t]ID_{RP}$.
Then, the IdP uses the user's permanent identity $u$ to calculate $PID_U = [u]PID_{RP} = [ut]ID_{RP}$,
similar to token signing by the token server in PrivacyPass/TrustToken.
IdP untraceability in UPPRESSO, i.e., the IdP cannot link $ID_{RP}$ and $[t]ID_{RP}$,
 roughly corresponds to unlinkability of token signing-redemption, i.e., the token server cannot link $T$ and $[e]T$  in PrivacyPass/TrustToken.

UPPRESSO leverages this cryptographic technique in very different ways.\footnote{We designed and implemented UPPRESSO in 2020 \cite{uppresso-arxiv}, and in 2023 someone reminded us that the identity transformations in UPPRESSO are very similar to the Hash(EC)DH OPRF.} 
First, 
it works among three parties, unlike the two-party PrivacyPass/TrustToken protocols.
UPPRESSO shares the user-selected random number $t$ with the target RP,
enabling it to derive the user's permanent account, i.e., $Acct = [t^{-1}]PID_{U}$. This account is ``obliviously'' determined by the IdP when it calculates the user's pseudo-identity.
Second, $u$ and $ID_{RP}$, which roughly correspond to $k$ and $T$ in PrivacyPass/TrustToken, are assigned as the permanent identities of the user and the RP, respectively, to establish the relationship among identities and accounts.
This is different from existing OPRF-based systems  \cite{privacypass, trusttoken, strong-oprf, oprf-bitcoin-wallet, pesto, oprf-ot-si, pp-ss, Private-Contact-Discovery, o-kms, oprf-deduplication} that always use $k$ as a server's secret key.
Finally, UPPRESSO leverages randomness of Hash(EC)DH OPRFs to ensure RP unlinkability, but this property is not utilized in PrivacyPass/TrustToken.
This ensures colluding RPs cannot link any logins across RPs,
even by sharing their knowledge about the pseudo-identities and permanent accounts in UPPRESSO.
This property offers desirable indistinguishability of different users visiting colluding RPs.
It roughly corresponds to the indistinguishability of different private keys for signing anonymous tokens, not considered in PrivacyPass/TrustToken. 

Although UPPRESSO mathematically employs the same technique in the identity transformations as the Hash(EC)DH OPRF \cite{oprf-proved,voprf-proved}, we require more properties of these algorithms. 
UPPRESSO essentially 
depends on the deterministicness property of pseudo-random functions to correctly derive the account for any $t$, obliviousness to ensure IdP untraceability, 
and pseudo-randomness to ensure RP unlinkability. 
In contrast, PrivacyPass and TrustToken \cite{privacypass,trusttoken} depend on only the properties of deterministicness and obliviousness.
In fact, UPPRESSO requires additional properties for secure SSO services 
    \cite{understanding-arxiv} (i.e., RP designation and user identification; see Theorems \ref{thm-u-id} and \ref{thm-rp-designation} in Section \ref{analysis-security}),
    which are not discussed in OPRFs \cite{sok-oprf,oprf-proved,voprf-proved}.
Thus, an OPRF is not always ready to work as the identity transformations in SSO.

%

\section{The Identity-Transformation Approach}
\label{sec:challenge}

In this section,
    we list the security requirements of SSO,
        and present the identity-transformation approach.

\subsection{Security Requirements for SSO Services}
\label{subsec:basicrequirements}

Non-anonymous SSO services are designed to allow a user to log into an honest RP as her unique account at this RP, 
by presenting identity tokens issued by a trusted IdP   \cite{OpenIDConnect,rfc6749,SAML,SAMLIdentifier,NIST2017draft}.
This security goal is achieved through the following properties of identity tokens.

First of all, \emph{authenticity}, \emph{confidentiality}, and \emph{integrity} of identity tokens are necessary to prevent forging, eavesdropping and tampering.
In SSO services, identity tokens are signed by the trusted IdP,
    transmitted over HTTPS/TLS \cite{OpenIDConnect, rfc6749, SAML},
    and finally forwarded to the target RPs by the authenticated user.
Common security mechanisms within a browser (e.g., the mechanisms of HTTP session and \verb+postMessage+ targetOrigin)
are also required for confidentiality of tokens \cite{GoogleIdIntegrate,de2014oauth,FettKS14,BrowserID} (see Section \ref{sec:web-design} for details).

Meanwhile, in each login the IdP issues an identity token that (\emph{a}) specifies the target RP (i.e., \emph{RP designation}) and (\emph{b}) identifies the authenticated user who requests this token (i.e., \emph{user identification}),
        as the enclosed (pseudo-)identities of RP and user \cite{OpenIDConnect,rfc6749,SAML} (see Section \ref{analysis-security} for details).
RP designation prevents malicious RPs from replaying received identity tokens to gain illegal access to other honest RPs as victim users.
So an RP usually checks the RP (pseudo-)identity enclosed in a token before accepting it 
\cite{OpenIDConnect,BrowserID,SPRESSO,NIST2017draft,POIDC,save-flow,up-sso,miso}
 and then allows the token holder to log in as the specified (pseudo-)identity (i.e., account).


\subsection{Identity Transformations in SSO}
\label{subsec:solutions}


Existing privacy-preserving solutions of SSO hide only RP identities or user identities
    (i.e., only IdP-based login tracing or RP-based identity linkage is prevented),
        or introduce an extra fully-trusted component to process the (pseudo-)identities of both RP and user before an identity token is issued.

UPPRESSO \emph{simultaneously} transforms the identities of RP and user in each login,
        without extra trusted components more than the honest-but-curious IdP,
         to provide untraceable and unlinkable SSO service.
%
Table \ref{tbl:notations-dilemma} lists the notations of the identity-transformation approach,
and a sub/super-script (i.e., $i$, $j$, or $l$) may be omitted if it does not cause ambiguity.

\begin{table}[t]
\footnotesize
    \caption{The (pseudo-)identities in the identity transformations}
    \centering
    \begin{tabular}{|l|p{5.15cm}|l|} \hline
    {\textbf{Notation}} & {\textbf{Definition}} & {\textbf{Lifecycle}} \\ \hline
    {$ID_{U_i}$} & {The $i$-th user's unique identity at the IdP.} & {Permanent} \\ \hline
    {$ID_{RP_j}$} & {The $j$-th RP's unique identity at the IdP.} & {Permanent} \\ \hline
    {$Acct_{i,j}$} & {The $i$-th user's account at the $j$-th RP.} & {Permanent} \\ \hline
    {$PID_{U_i,j}^l$} & {The $i$-th user's pseudo-identity in her $l$-th login visiting the $j$-th RP.} & {Ephemeral} \\ \hline
    {$PID_{RP_j}^l$} & {The $j$-th RP's pseudo-identity in the user's $l$-th login visiting this RP.} & {Ephemeral} \\ \hline
    \end{tabular}
    \label{tbl:notations-dilemma}
\end{table}



First of all, each registered user owns her account unique at an RP.
$\mathcal{F}_{Acct\ast}(ID_{U_i}, ID_{RP_j}) = Acct_{i,j}$ is proposed to determine a user's account at an RP.
An account is \emph{meaningful} at an RP if it belongs to some \emph{registered} user at this RP,
while a \emph{meaningless} account $A_m$ at $RP_j$ does not belong to any registered user, i.e., $\forall ID_{U_i} \in \mathbbm{ID}_U = \{ID_{U_{i; 1 \leq i \leq s}}\}: A_m \neq \mathcal{F}_{Acct\ast}(ID_{U_i}, ID_{RP_j})$.
That is, meaningless accounts belongs to ``potential'' users but not registered yet.

To ensure IdP untraceability (or prevent IdP-based login tracing),
    a user initiates a login by calculating an \emph{ephemeral} pseudo-identity $PID^l_{RP_j}$  for the target RP and sending an identity-token request for $PID^l_{RP_j}$ (but not $ID_{RP_j}$) to an IdP.

After authenticating the initiating user as $ID_{U_i}$, the IdP calculates an \emph{ephemeral} $PID^l_{U_i,j}$ based on $ID_{U_i}$ and $PID^l_{RP_j}$, and then issues an identity token that binds $PID^l_{U_i,j}$ and $PID^l_{RP_j}$.
To ensure RP unlinkability (i.e., prevent RP-based identity linkage),
    $ID_{U_i}$ is not enclosed in the identity token.

On receiving a signed token, the RP calculates the user's \emph{permanent} account (i.e., $Acct_{i,j}$) based on $PID^l_{U_i,j}$ and $PID^l_{RP_j}$,
and allows the token holder to log in as $Acct_{i,j}$.

Therefore, in addition to $\mathcal{F}_{Acct\ast}(ID_{U}, ID_{RP}) = Acct$,
we propose the following identity transformations in a login initiated by a user to visit an RP in privacy-preserving SSO:
\begin{itemize}
\setlength{\topsep}{0pt}
\setlength{\partopsep}{0pt}
\setlength{\itemsep}{0pt}
\setlength{\parsep}{0pt}
\setlength{\parskip}{0pt}

\item
$\mathcal{F}_{PID_{RP}}(ID_{RP}) = PID_{RP}$, calculated by the initiating user,
    where $ID_{RP}$ is the target RP's identity.
In the IdP's view,
$\mathcal{F}_{PID_{RP}}()$ is a one-way function and $PID_{RP}$
is \emph{indistinguishable} from random variables.
\item
$\mathcal{F}_{PID_U}(ID_U, PID_{RP}) = PID_{U}$, calculated by the IdP,
    where $ID_U$ identifies the user requesting tokens.
In the target RP's view,
    $\mathcal{F}_{PID_U}()$ is a one-way function and $PID_{U}$ is \emph{indistinguishable} from random variables.
\item
$\mathcal{F}_{Acct}(PID_{U}, PID_{RP}) = Acct$, calculated by the target RP.
In the user's multiple logins visiting the RP,
    $Acct = \mathcal{F}_{Acct\ast}(ID_{U}, ID_{RP})$ is always derived.
\end{itemize}

\begin{figure}[tb]
  \centering
  \includegraphics[width=1.00\linewidth]{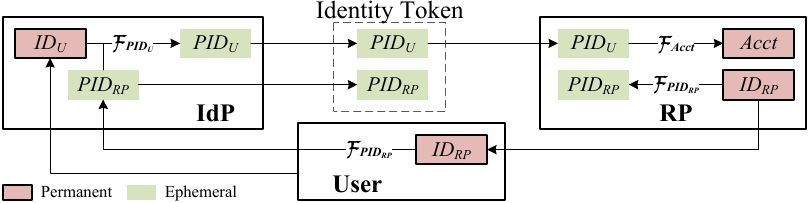}
  \caption{Identity transformations in UPPRESSO} 
  \label{fig:IDCorrelation}
\end{figure}

The relationships among the (pseudo-)identities are depicted in Figure \ref{fig:IDCorrelation}.
Red and green blocks represent \emph{permanent} identities and \emph{ephemeral} pseudo-identities, respectively, and 
labeled arrows denote the transformations of (pseudo-)identities.
$PID_{RP}$ and $PID_{U}$ are ephemeral;
    otherwise, the IdP could learn the visited RP.
$PID_{U,j}$ for visiting different RPs are independent of each other;
    otherwise, colluding RPs could learn something on $ID_U$ from $PID_{U,j}$s and link the logins.

In privacy-preserving SSO services integrating the identity transformations,
    an identity token \emph{explicitly binds} $PID_{RP_j} = \mathcal{F}_{PID_{RP}}(ID_{RP_j})$ but \emph{implicitly designates} only the RP with $ID_{RP_j}$.
The following properties are required for the identity transformations:
\begin{itemize}
\setlength{\topsep}{0pt}
\setlength{\partopsep}{0pt}
\setlength{\itemsep}{0pt}
\setlength{\parsep}{0pt}
\setlength{\parskip}{0pt}

\item \textbf{Account Uniqueness.} Each user owns an account locally unique at an RP, determined by $\mathcal{F}_{Acct\ast}(ID_{U}, ID_{RP})$.
\item \textbf{RP Designation.} 
An identity token is rejected, at any honest RPs other than the designated one.

\item \textbf{User Identification.} From an identity token,
    the designated honest RP derives only the account exactly identifying the user requesting this token.
\item \textbf{IdP Untraceability.}
The IdP learns nothing on the visited RP from the identity-token requests.

\item \textbf{RP Unlinkability.}
Colluding RPs 
 cannot link any login initiated by an honest user visiting an RP,
  to any logins visiting any other colluding RPs by honest users.
\end{itemize}

In addition to the basic requirement of account uniqueness,
    RP designation and user identification ensure security of SSO services,
    while IdP untraceability and RP unlinkability protect user privacy
(see Section \ref{sec:analysis} for the formal proofs). 
\section{The Designs of UPPRESSO}
\label{sec:UPPRESSO}


\subsection{Threat Model}
\label{subsec:threatmodel}
The system consists of an honest-but-curious IdP, as well as several honest or malicious RPs and users. This model is in line with widely-deployed SSO services in the Internet \cite{OpenIDConnect,rfc6749, SAML, SAMLIdentifier}.

\noindent \textbf{Honest-but-curious IdP.} The IdP strictly follows the protocols,
 while remaining interested in learning about a user's login activities.
For example, it could store received messages to infer the relationship among $ID_U$, $ID_{RP}$, $PID_{U}$, and $PID_{RP}$.
It never actively violates the protocols,
 so a user-i script downloaded from the IdP also follows the protocols.

The IdP maintains its private key well for signing identity tokens and RP certificates, 
and adversaries cannot forge such signed messages.

\noindent \textbf{Malicious RPs.}
Adversaries could control a set of RPs by registering at the IdP as an RP or exploiting vulnerabilities to compromise some RPs.
Malicious RPs could behave arbitrarily, attempting to break the security or privacy guarantees for honest users and RPs \cite{SPRESSO, FettKS14,POIDC,miso}.

\noindent \textbf{Malicious users.} Adversaries could control a set of users by stealing their credentials or registering Sybil users in the system.
A malicious user could modify, insert, drop, or replay messages or even behave arbitrarily.

\noindent \textbf{Colluding RPs and users.}
Malicious users and RPs could collude,
 attempting to break the security or privacy guarantees;
that is,
their objective is to (\emph{a}) impersonate a victim user at honest RPs, (\emph{b}) entice an honest user to log into an honest RP under another user's account,
 and (\emph{c}) link an honest user's logins visiting colluding RPs.
For example, they could manipulate $PID_{RP}$ and $t$ in a login, attempting to (\emph{a}) entice honest users to return an identity token that could be accepted by some honest RP or (\emph{b}) affect the generation of $PID_U$ to analyze the relationship between $ID_U$ and $PID_U$.

\subsection{Assumptions}
We assume secure communications between honest entities, and the cryptographic primitives are secure. The software stack of an honest entity (e.g., a browser) is correctly implemented to transmit messages to receivers as expected.
These assumptions are realized by the commonly-used mechanisms in popular SSO services
 \cite{OpenIDConnect, rfc6749, SAML,GoogleIdIntegrate,de2014oauth,FettKS14,BrowserID,uber} (e.g., HTTPS/TLS, HTTP session maintenance, and the \verb+postMessage+ HTML5 API in COTS browsers; see Section \ref{sec:web-design} for specific designs for web applications).

As $PID_{RP}$ but not $ID_{RP}$ is processed by the IdP,
it also requires that no other information about the visited RP is leaked to the IdP,
 in the requesting, receiving and forwarding of identity tokens.
We realize this by specific designs in Section \ref{sec:web-design}.
Our work focuses on the privacy threats introduced by SSO protocols, and does not consider the tracking of user activities by network traffic analysis or crafted web pages, as they can be prevented by other defenses.

UPPRESSO is designed for users who value privacy,
        and provides an easy-to-deploy option for such users.
So privacy leakages due to re-identification by distinctive attributes across RPs are out of our scope.
A user never consents to enclose distinctive attributes, such as telephone number, Email address, etc., in tokens or sets such attributes at any RP.

\subsection{Identity-Transformation Algorithms}
\label{subsec:overview}

We propose the following identity transformations on an elliptic curve $\mathbb{E}$,
where $G$ is a generator on $\mathbb{E}$ of order $n$.

\noindent {\bf $\boldsymbol{ID_{\boldsymbol{U}}}$, $\boldsymbol{ID_{\boldsymbol{RP}}}$ and $\boldsymbol{Acct}$.}
The IdP assigns a unique random integer $u \in \mathbb{Z}_n$ to a registered user (i.e., $ID_U = u$),
  and randomly selects unique $ID_{RP} = [r]G$ for a registered RP. 
Here, 
 $[r]G$ denotes the addition of $G$ on the curve $r$ times.
So $Acct_{i,j} = \mathcal{F}_{Acct\ast}(ID_{U_i}, ID_{RP_j})= [ID_{U_i}]ID_{RP_j} =[u_ir_j]G$ is automatically assigned to a registered user at every RP.

All entities except the IdP
    know nothing about a user's identity $ID_U = u$; otherwise, colluding RPs could calculate $[u]ID_{RP_j}$ for any known $u$ and link these accounts.
This requires that $|\mathbb{Z}_n|$ is sufficiently large against exhaustive attacks; for example, $n$ is approximately $2^{256}$ for the NIST P256 elliptic curve and the amount of users is usually less than $2^{64}$.

$ID_{RP} = [r]G$ and $Acct = [ID_U]ID_{RP}$ are publicly-known,
 but $r$ is always kept secret;
otherwise, two colluding RPs with $ID_{RP_j} = [r]G$ and $ID_{RP_{j'}} = [r']G$ could link a user's accounts by checking whether $[r']Acct_j$ is equal to $[r]Acct_{j'}$ or not.

In UPPRESSO $r$ is processed only \emph{once} by the IdP,
    while $u$ is used only by the IdP \emph{internally} and
 not enclosed in any message; see Section \ref{implementations} for details.

\noindent {\bf $\boldsymbol{ID_{\boldsymbol{RP}}}$-$\boldsymbol{PID_{\boldsymbol{RP}}}$ Transformation.} When visiting an RP,
a user selects a random number $t \in \mathbb{Z}_n$
    and calculates $PID_{RP}$ as below. She also sends $t$ to the RP.
\begin{equation}
PID_{RP} = \mathcal{F}_{PID_{RP}}(ID_{RP},t) = [t]{ID_{RP}}
\label{equ:PIDRP}
\end{equation}



\noindent {\bf $\boldsymbol{ID_U}$-$\boldsymbol{PID_U}$ Transformation.}
After receiving an identity-token request for $PID_{RP}$ from a user authenticated as $ID_U$, the IdP calculates $PID_{U}$ as below.
\begin{equation}
PID_{U} = \mathcal{F}_{PID_U}(ID_U, PID_{RP}) =
  [{ID_U}]{PID_{RP}}
 \label{equ:PIDU}
\end{equation}


\noindent {\bf $\boldsymbol{PID_U}$-$\boldsymbol{Acct}$ Transformation.}
After verifying a signed identity token, the RP calculates $Acct$ as follows.
\begin{equation}
Acct = \mathcal{F}_{Acct}(PID_{U},t)
   = [t^{-1} \bmod n]PID_{U}
   \label{equ:Account}
\end{equation}

First of all, the above designs ensure \emph{account uniqueness}:
    At an RP, every user owns a unique account, because $ID_{RP} = [r]G$ is also a generator of $\mathbb{E}$ and then
     $Acct = \mathcal{F}_{Acct\ast}()=[ur]G$ is unique locally at this RP.

In a login involving no adversary,
 the visited RP always derives the account belonging to the user who requests the identity token,
    by calculating $\mathcal{F}_{Acct}(PID_{U}, PID_{RP})$.
From Equations \ref{equ:PIDRP}, \ref{equ:PIDU}, and \ref{equ:Account}, it is derived that
\begin{equation}
   Acct =  [t^{-1}utr]G = [ur]G = \mathcal{F}_{Acct\ast}(ID_U, ID_{RP})
   \label{equ:AccountNotChanged}
\end{equation}

In Section \ref{sec:analysis}
we formally prove the other required properties (i.e.,
    \emph{RP designation}, \emph{user identification}, \emph{IdP untraceability}, and \emph{RP unlinkability}) of the identity transformations under the adversarial scenarios.

\subsection{The Designs Specific for Web Applications}
\label{sec:web-design}

When the required properties of identity transformations are satisfied,
    in UPPRESSO we need to additionally ensure:
\begin{itemize}
  \item[1.] $PID_{RP} = [t]{ID_{RP}}$ is calculated based on the target RP's identity but not any other RPs' (i.e., the target RP is designated),
  \item[2.]
  An identity token is forwarded to only the target RP web server in a browser,
    as authenticity and integrity of tokens have been ensured by the IdP's signatures
        and they are transmitted over HTTPS/TLS, 
\item[3.] 
$t$ is known only to the target RP and the initiating user (but not the IdP; otherwise, the IdP is able to infer $ID_{RP}$ from $PID_{RP}$),
and no direct privacy leakage in the requesting of identity tokens
(e.g., $ReEnptUrl_{RP}$ or $ID_{RP}$ is not included in the requests,
and no HTTP \texttt{referer} header about the RP web server is leaked in HTTP requests to the IdP).
\end{itemize}

In an SSO system implementing OIDC with pop-up UX, as shown in Figure \ref{fig:OpenID},
    a user requests a token by sending the target RP's identity (i.e., $ID_{RP}$) via the user-i window,
    and the IdP replies with the signed token and also the RP's URL to receive tokens (i.e., $ReEnptUrl_{RP}$).
Then, after receiving the token, the user-i script forwards it to the user-r script by \verb+postMessage+
    which is restricted by the origin of $ReEnptUrl_{RP}$ \cite{SPRESSO,MITREid,BrowserID,GoogleIdIntegrate,de2014oauth,OpenIDConnect}.
The \verb+postMessage+ targetOrigin mechanism \cite{postm-targeto} restricts the recipient,
    so the token is forwarded to exactly the target RP because (\emph{a}) only scripts downloaded from the origin of $ReEnptUrl_{RP}$ are legitimate recipients of the forwarded tokens
    and (\emph{b}) the relationship between $ReEnptUrl_{RP}$ and $ID_{RP}$ is maintained at the honest IdP server.

\begin{figure*}[htb]
  \centering
  \includegraphics[width=0.79\linewidth]{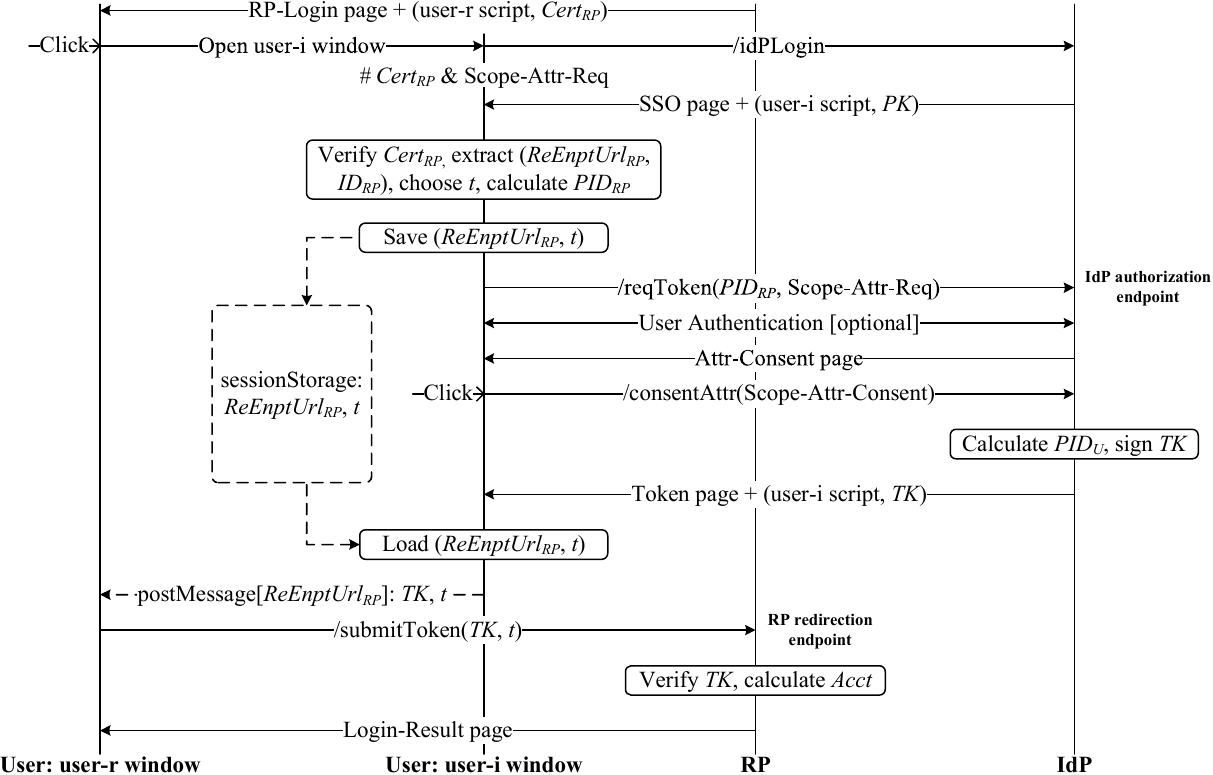}
  \caption{The SSO login flow of UPPRESSO}
  \label{fig:process}
\end{figure*}

The following improvements are proposed  in UPPRESSO,
    to satisfy the additional requirements in an OIDC-compatible way.
First of all, the IdP receives $PID_{RP}$ (but not $ID_{RP}$) in an identity-token request,
    so it cannot reply with $ReEnptUrl_{RP}$ to the user-i window.
An RP certificate, denoted as $Cert_{RP}$, is introduced to bind $ReEnptUrl_{RP}$ and $ID_{RP}$, signed by the IdP during the RP's initial registration.
%
%
%
Then, in each login, 
as shown in Figure \ref{fig:process},
when the user-i window is popped up,
    $Cert_{RP}$ and the scope of request user attributes are processed with the HTTP request to the IdP,
    all of which are generated by the user-r script downloaded from the target RP.
This HTTP request downloads the user-i script,
    which in turn verifies $Cert_{RP}$ to extract $ID_{RP}$ and $ReEnptUrl_{RP}$, and calculates $PID_{RP} = [t]ID_{RP}$. 
Then, after the IdP issues a token,
    the user-i script forwards it to the user-r script by \verb+postMessage+
    which is restricted by the origin of $ReEnptUrl_{RP}$.
Besides, on receiving an identity-token request, the IdP web server checks the included HTTP \texttt{origin} header to ensure it is sent by the user-i script,
which will honestly verify $Cert_{RP}$ to ensure the correct relationship between $ReEnptUrl_{RP}$ and $ID_{RP}$.
This ensures that $PID_{RP} = [t]{ID_{RP}}$ is calculated based on the visited RP's identity
    and only the designated RP receives the token,
    because (\emph{a}) $ReEnptUrl_{RP}$ and $ID_{RP}$ are signed in $Cert_{RP}$
        and (\emph{b}) only scripts from the origin of $ReEnptUrl_{RP}$ are legitimate recipients of \verb+postMessage+.

In the HTTP request to pop up the user-i window, $Cert_{RP}$ and the scope of requested user attributes are carried and processed as URL \emph{fragments} (but not query parameters) \cite{url-fragment}, which are identified by \texttt{\#} instead of \texttt{?},
    so that they are not sent to the IdP web server but locally processed by the browser.
Meanwhile, we save $t$ and $ReEnptUrl_{RP}$ using the \verb+sessionStorage+ HTML object \cite{htmlStorageHtml},
    because the user-i window may be refreshed to finish user authentication and attribute consent.
That is, $t$, $ReEnptUrl_{RP}$ and $Cert_{RP}$ are processed within browsers,
    not sent to the IdP.

To prevent referer leakage in the HTTP request to pop up the user-i window,
 we need to ensure this request does not carry an HTTP \texttt{referer} header, which reveals the visited RP's domain to the IdP web server.\footnote{In original OIDC services this privacy leakage commonly exists, but does not matter because such services do not aim to prevent IdP-based login tracing.} 
%
It is achieved by setting
\verb+<meta name="referrer" content="no-referrer">+
in the RP's SSO login page \cite{referer_policy}.
Finally, 
two endpoints 
    specified in OIDC \cite{rfc6749,OpenIDConnect} for the implicit login flow are implemented in an OIDC-compatible way.
When receiving a token request for an ``unregistered'' $ID_{RP}$ (i.e., ephemeral $PID_{RP}$),
the IdP issues an identity token to an ``empty'' $ReEnptUrl_{RP}$,
    and the calculation of $PID_{U}$ is considered as a special way to assign PPIDs.
The RP's redirection endpoint is also implemented compatibly,
    where $t$ is sent along with the tokens as an optional parameter.
Moreover, the IdP's public key is set in the user-i script to verify $Cert_{RP}$,
so a user configures nothing locally,
 like in other OIDC implementations \cite{OpenIDConnect, GoogleIdIntegrate,de2014oauth,uber}.

\subsection{The UPPRESSO protocol}
\label{implementations}

\noindent \textbf{System Initialization.}
$\mathbb{E}$, $G$ and $n$ are set up and publicly published.
An IdP generates a key pair ($SK$, $PK$) to sign and verify identity tokens and RP certificates.

\noindent\textbf{RP Registration.}
Each RP registers itself at the IdP to obtain $ID_{RP}$ and its RP certificate $Cert_{RP}$ as follows.
This may be conducted by face-to-face or online \cite{OpenID-dy-regi} means.
\begin{enumerate}
\setlength{\topsep}{0pt}
\setlength{\partopsep}{0pt}
\setlength{\itemsep}{0pt}
\setlength{\parsep}{0pt}
\setlength{\parskip}{0pt}
\item[1.]
An RP pre-installs $PK$ by trusted means.
It submits a registration request, including the URL to receive identity tokens (i.e., $ReEnptUrl_{RP}$) and other information.
\item[2.]
After examining the request,
the IdP randomly selects $r \in \mathbb{Z}_n$
        and assigns a \emph{unique} point $[r]G$ to the RP as its identity.
Note that $r$ is not processed any more and then known to nobody
 due to the elliptic curve discrete logarithm problem (ECDLP).
The IdP then signs $Cert_{RP} = [ID_{RP}, ReEnptUrl_{RP}, *]_{SK}$,
     where $[\cdot]_{SK}$ is a message signed using $SK$ and $*$ is supplementary information such as the RP's unambiguous name.
\item[3.]
The RP verifies $Cert_{RP}$ using $PK$, and accepts $ID_{RP}$ and $Cert_{RP}$.
\end{enumerate}

\noindent\textbf{User Registration.}
Each user sets up her unique username and the corresponding credential for the IdP.
The IdP assigns
a random but unique identity $ID_U = u\in \mathbb{Z}_n$ to the user.

$ID_U$ is known \emph{only} to the IdP
    and other entities know nothing about $ID_U$.
$ID_U$ is used only by the IdP \emph{internally} in UPPRESSO,
 not enclosed in any message.
For example, a user's identity is generated and always restored by hashing her username concatenated with the IdP's private key.
Then,
 $ID_U$ is never stored in hard disks,
 and protected almost the same as the IdP's private key
because it is \emph{only} used to calculate $PID_{U}$ as the IdP is signing an identity token binding $PID_{U}=
  [{ID_U}]{PID_{RP}}$.

\noindent\textbf{SSO Login.} 
A login flow 
involves three steps: Identity-token requesting, generation, and acceptance.
In Figure \ref{fig:process}, the IdP's and RP's operations are connected by two vertical lines, respectively. The user operations are split into two groups in different browser windows by vertical lines, one communicating with the IdP and the other with the RP. Solid horizontal lines indicate messages exchanged between a user and the IdP (or the RP), while dotted lines represent the mechanisms within a browser:
(\emph{a}) \verb+postMessage+ invocations between two scripts (or browser windows)
or (\emph{b}) \verb+sessionStorage+ objects accessed by refreshed pages of an HTTP session.

\noindent 1. {\em Identity-Token Requesting.}
The user requests an identity token for $PID_{RP} = [t]{ID_{RP}}$.
\begin{itemize}
\setlength{\topsep}{0pt}
\setlength{\partopsep}{0pt}
\setlength{\itemsep}{0pt}
\setlength{\parsep}{0pt}
\setlength{\parskip}{0pt}
\item[1.1]
The target RP provides an SSO login page with the user-r script,
    and $Cert_{RP}$ is set in this script.
The user clicks the SSO login button,
    to pop up another browser window responsible for communicating with the IdP.
\item[1.2] The user-i window downloads the user-i script,
which processes $Cert_{RP}$ and the scope of requested attributes sent from the user-r window.
It verifies $Cert_{RP}$, extracts $ID_{RP}$ and $ReEnptUrl_{RP}$ from $Cert_{RP}$, 
chooses a random number $t \in \mathbb{Z}_n$,
    and calculates $PID_{RP}=[t]{ID_{RP}}$.
\item[1.3] 
Then, the user-i script saves $ReEnptUrl_{RP}$ and $t$ in the \verb+sessionStorage+ HTML5 object,
    and requests an identity token from the IdP by sending $PID_{RP}$ and the scope of requested attributes.
\end{itemize}

\noindent 2. {\em Identity-Token Generation.}
The IdP calculates $PID_U = [ID_U]{PID_{RP}}$ and signs an identity token as follows. 
\begin{itemize}
\setlength{\topsep}{0pt}
\setlength{\partopsep}{0pt}
\setlength{\itemsep}{0pt}
\setlength{\parsep}{0pt}
\setlength{\parskip}{0pt}
\item[2.1]
On receiving an identity-token request,
the IdP authenticates the initiating user as $ID_U$, if not authenticated yet.
Then, it obtains the user's consent to enclose attributes.
\item[2.2]
The IdP calculates $PID_U = [ID_U]{PID_{RP}}$, and signs $TK = [PID_{RP}, PID_U, Iss, Validity, Attr]_{SK}$,
where $Iss$ identifies the IdP, $Validity$ indicates the validity period, and $Attr$ contains the user attributes.
\item[2.3] The IdP replies with the identity token to the user-i window,
    as well as the user-i script to forward this token.
\end{itemize}

\noindent 3. {\em Identity-Token Acceptance.}
The RP receives the identity token and allows the user to log in.
\begin{itemize}
\setlength{\topsep}{0pt}
\setlength{\partopsep}{0pt}
\setlength{\itemsep}{0pt}
\setlength{\parsep}{0pt}
\setlength{\parskip}{0pt}
\item [3.1]
From the \verb+sessionStorage+ object
the user-i script loads $ReEnptUrl_{RP}$ and $t$, 
    and forwards the identity token and $t$ to the user-r window by \verb+postMessage+ which is restricted by the origin of $ReEnptUrl_{RP}$.
\item[3.2] The user-r script submits it to the RP.
The RP verifies the signature and the validity period of the received token, 
and then calculates $Acct = [t^{-1}]{PID_U}$.
Only if $Acct$ is meaningful, the RP allows the user to log in.

\end{itemize}

If any verification fails, the flow will be immediately terminated.
For instance, when receiving invalid $Cert_{RP}$ the user halts it.
The IdP rejects an identity-token request if the received $PID_{RP}$ is not a point on $\mathbb{E}$, and the RP rejects a token if it is not well signed or a meaningless account is derived.

In Step 3.2 an RP does not check whether $PID_{RP}$ in the token is equal to $[t]ID_{RP}$ or not,
    and this efficient design does not result in attacks, which is proved in Section \ref{analysis-security}.

In order to distinguish meaningless accounts from meaningful ones in Step 3.2,
it requires that the RP \emph{locally} maintains an up-to-date account set for all registered users,
i.e., 
$\{Acct_i\}_j=\{[u_{i; 1 \leq i \leq s} r_j]G\}$ and $|\{Acct_i\}_j|=s$.
We discuss the means of account synchronization in Section \ref{account-syn}.

\section{Security and Privacy Analysis}
\label{sec:analysis}

As account uniqueness is analyzed in Section \ref{subsec:overview}
this section proves security (i.e., RP designation and user identification) and privacy (i.e., IdP untraceability and RP unlinkability) of the identity transformations in UPPRESSO.

\subsection{Adversarial Scenarios}

According to our design goals 
 and the potential adversaries listed in Section \ref{subsec:threatmodel}, we consider three adversarial scenarios as below
 and the security and privacy guarantees are proved against different adversaries.

First of all, security (i.e., RP designation and user identification) is proved with an honest IdP but against \emph{malicious RPs and users}.
In the proofs, malicious entities could collude \cite{FettKS14,BrowserID,SPRESSO},
  attempting to (\emph{a}) impersonate an honest user to log into some honest RP
   or (\emph{b}) entice an honest user to log into an honest RP under another's account, 
by arbitrarily manipulating $t$ and/or sending $TK$ to an RP not designated.

Provided that the security properties are satisfied,
    we analyze the privacy properties \emph{only} for successful logins.
It makes no sense to track user activities
     from malicious behaviors deviating from the protocols, resulting unsuccessful logins.
So no malicious entities are considered in the proofs of privacy.
In particular, (\emph{a}) IdP untraceability is proved in Section \ref{subsec:IdP-privacy} against \emph{an honest-but-curious IdP}, which tries to infer the identities of the \emph{honest} RPs being visited by an \emph{honest} user,
and (\emph{b}) RP unlinkability is analyzed in Section \ref{subsec:RP-privacy} against \emph{a group of semi-honest RPs and users} which follow the protocols but share their information,
    attempting to link an honest user's accounts across these semi-honest RPs.

\subsection{Security}
\label{analysis-security}

In secure SSO systems, an identity token denoted as $TK$, which is requested by a user to visit an RP,
    \emph{enables only this user to log into only the honest target RP as her account at this RP} (i.e., RP designation and user identification). It makes no sense to discuss the login results at malicious RPs.

Let's assume totally $s$ users and $p$ RPs in UPPRESSO,
    whose identities are denoted as $\mathbbm{ID}_U = \mathbbm{u} = \{u_{i; 1 \leq i \leq s}\}$ and $\mathbbm{ID}_{RP} = \{[r_{j;1 \leq j \leq p}]G\}$, respectively.
$Acct_{i,j}$ and $ID_{RP_j}$ are publicly-known, while $u_{i}$ and $r_{j}$ are kept secret to adversaries.

$TK$ is signed by the IdP to bind $PID_{RP}$ and $PID_U$,
where (\emph{a}) the RP with $ID_{RP}$ is designated if $PID_{RP} = [t]ID_{RP}$ is calculated
    and (\emph{b}) $PID_U = [ID_U]PID_{RP}$ is calculated based on the initiating user's identity $ID_U$.

RP designation and user identification are proved in Theorems \ref{thm-rp-designation} and \ref{thm-u-id}, respectively, against malicious RPs colluding with users.
Besides, as mentioned in Section \ref{implementations}, 
in Step 3.2 an (honest) RP directly derives accounts from any signed and unexpired identity token, without checking whether $PID_{RP}$ in the token is equal to $[t]ID_{RP}$ or not.
So in the proofs we do not assume any relationship among $t$, $ID_{RP}$ and $PID_{RP}$ which are received by an RP.




RP designation means that, \emph{from $TK$ only the designated honest RP derives meaningful accounts}.
We prove this by showing that,
    at any honest RPs other than the designated one, no meaningful account will be derived from $TK$.


\vspace{1.5mm}
\begin{thm}[RP Designation]
When given known $\mathbbm{ID}_{RP}$, known $\mathbbm{A}cct$ and unknown $\mathbbm{ID}_U$ defined as above,
  malicious adversaries cannot find ${j_1}$, ${j_2}$, ${i_1}$, ${i_2}$, $t_1$, and $t_2$ which satisfy that $\mathcal{F}_{Acct}(\mathcal{F}_{PID_U}(ID_{U_{i_1}}, PID_{RP}^{*}), t_2) = \mathcal{F}_{Acct\ast}(ID_{U_{i_2}}, ID_{RP_{j_2}})$,
where $PID_{RP}^{*}$ may be calculated as $\mathcal{F}_{PID_{RP}}(ID_{RP_{j_1}}, t_1)$
    (i.e., $RP_{j_1}$ is designated)
    or even arbitrarily generated (i.e., no RP is designated),
    ${j_1} \neq {j_2}$, $1 \leq j_1, j_2 \leq p$, and $1 \leq i_1, i_2 \leq s$.
\label{thm-rp-designation}
\end{thm}

\noindent\textbf{\textsc{Proof.}} 
If adversaries could find ${j_1}$, ${j_2}$, ${i_1}$, ${i_2}$, $t_1$ and $t_2$ which satisfy that $[{t_2}^{-1}t_1ID_{U_{i_1}}]ID_{RP_{j_1}} = [ID_{U_{i_2}}]ID_{RP_{j_2}}$
 where ${j_1} \neq {j_2}$,
 the ECDLP of ${t_2}^{-1}t_1 = \log_{[{u_{i_1}}{r_{j_1}}]G}{[{u_{i_2}}{r_{j_2}}]G}
 = \log_{Acct_{i_1,j_1}}{Acct_{i_2,j_2}}$ would be solved.

The adversary might alternatively attempt to find $PID_{RP}^{*}$, ${j_2}$, ${i_1}$, ${i_2}$ and $t_2$ which satisfy that $[{{t_2}^{-1}}ID_{U_{i_1}}]PID_{RP}^{*} = [ID_{U_{i_2}}]ID_{RP_{j_2}}$,
        but this cannot be solved due to unknown $ID_{U_{i}}$ when ${i_1} \neq {i_2}$.
Or, if ${i_1} = {i_2}$, 
    it would require the adversary to solve $[{{t_2}^{-1}}]PID_{RP}^{*} = ID_{RP_{j_2}}$
        or $PID_{RP}^{*} = [t_2]ID_{RP_{j_2}}$.
That is,
    the adversary chooses $t_2$, calculates $PID_{RP}^{*} = [t_2]ID_{RP_{j_2}}$, requests a token for $PID_{RP}^{*}$, and then logs into $RP_{j_2}$ to ``impersonate'' himself.
This is not an attack: $RP_{j_2}$ is actually designated by $TK$ and the adversary's account is derived at $RP_{j_2}$.
\hfill $\square$

\vspace{1.5mm}

User identification means that,
\emph{from $TK$ the designated honest RP derives only the meaningful account belonging to the user requesting $TK$}.
We prove this property by showing that,
    at the RP designated by any $TK$,
        a meaningless account not belonging to the initiating user will never be derived.

\vspace{1.5mm}
\begin{thm}[User Identification] Given unknown $\mathbbm{ID}_U$ and known $\mathbbm{A}cct$,
for any known $ID_{RP} \in \mathbbm{ID}_{RP}$, 
     malicious adversaries cannot find $\check{t}$ and $TK$ binding $PID_{RP} = \mathcal{F}_{PID_{RP}}(ID_{RP}, \hat{t})$
     and $PID_{\hat{U}}=\mathcal{F}_{PID_U}(ID_{\hat{U}}, PID_{RP})$ satisfying that $\mathcal{F}_{Acct}(PID_{\hat{U}}, \check{t}) = \mathcal{F}_{Acct\ast}(ID_{\check{U}}, ID_{RP})$, where $ID_{\hat{U}} \neq ID_{\check{U}}$ and $ID_{\hat{U}},ID_{\check{U}} \in \mathbbm{ID}_U$.
\label{thm-u-id}
\end{thm}

\noindent\textbf{\textsc{Proof.}}
It requires that, for any $ID_{RP}$, adversaries cannot find $\hat{t}$ and $\check{t}$ satisfying that $\mathcal{F}_{Acct}(PID_{\hat{U}}, \check{t}) = [\check{t}^{-1}]PID_{\hat{U}} = [\check{t}^{-1}ID_{\hat{U}}]PID_{RP}
= [\check{t}^{-1}\hat{t}ID_{\hat{U}}]ID_{RP}
    = [ID_{\check{U}}]ID_{RP} = \mathcal{F}_{Acct\ast}(ID_{\check{U}}, ID_{RP})$,
where $ID_{\hat{U}} \neq ID_{\check{U}}$.

When $u_{i}$ and $r_{j}$ are kept unknown, this is equivalent to the elliptic-curve discrete logarithm problem (ECDLP) of $\check{t}^{-1}\hat{t} = \log_{[\hat{u}r]G}{[\check{u}r]G} = \log_{\hat{A}cct}{\check{A}cct}$.
\hfill $\square$

\subsection{Privacy against IdP-based Login Tracing}
\label{subsec:IdP-privacy}

The honest IdP would attempt to infer the visited RPs,
by collecting information from the identity-token requests.
According to the designs and the detailed protocol in Sections \ref{sec:web-design} and \ref{implementations},
 it does not obtain any information about the target RP in a login (e.g., $ID_{RP}$, $ReEnptUrl_{RP}$ or $Cert_{RP}$), except the RP's ephemeral pseudo-identity $PID_{RP}$.

We prove that, \emph{$PID_{RP}$ is indistinguishable from a random variable on $\mathbb{E}$ to the IdP},
    and then
 it cannot (\emph{a}) link multiple logins visiting an RP or (\emph{b}) distinguish a login initiated by a user to visit any RP from those logins visiting other RPs by this user. 
Therefore, IdP untraceability is ensured.


\vspace{1.5mm}
\begin{thm}[IdP Untraceability]
\emph{The honest-but-curious IdP cannot distinguish $PID_{RP} = [t]ID_{RP}$ from a random variable on $\mathbb{E}$, where $t$ is random in $\mathbb{Z}_n$ and kept unknown to the IdP.}\label{thm-idp-untraceability}
\end{thm}

\noindent\textbf{\textsc{Proof.}}
As $G$ is a generator on $\mathbb{E}$ of order $n$, $ID_{RP} = [r]G$ is also a generator of order $n$.
As $t$ is uniformly-random in $\mathbb{Z}_n$ and kept unknown, $PID_{RP} = [t]ID_{RP}$ is \emph{indistinguishable} from a point that is uniformly-random on $\mathbb{E}$.
\hfill $\square$

\vspace{1.5mm}

\noarxiv{This property actually has been proved in the Hash(EC)DH OPRF \cite{oprf-proved,voprf-proved} as obliviousness, where the OPRF server works similarly to the IdP and learns nothing about a client's inputs or outputs of the evaluated pseudo-random function (i.e., $ID_{RP}$ or $Acct$ in UPPRESSO).}

\subsection{Privacy against RP-based Identity Linkage}
\label{subsec:RP-privacy}

Semi-honest RPs and users share their information, attempting to 
link an honest user's accounts across the semi-honest RPs.
In each login, an RP obtains \emph{only} a random number $t$ and an identity token enclosing $PID_{RP}$ and $PID_U$. From the token, it learns only an ephemeral pseudo-identity $PID_{U} = [{ID_U}]{PID_{RP}}$ of the initiating user, from which it derives a \emph{permanent} locally-unique identifier (or account) $Acct = [ID_U]ID_{RP}$.
It cannot directly calculate $ID_U$ from $PID_{U}$ or $Acct$ due to the ECDLP.

Given the trapdoor $t$, 
    adversaries can transform $PID_{RP}$ and $PID_U$ into $ID_{RP}$ and $Acct$, respectively, and vice versa.
So we denote all the information that an RP learns in a login as a tuple $L =(ID_{RP}, t, Acct)=([r]G, t, [ur]G)$.

When $c$ RPs share information with each other, they create a shared view of all logins visiting any of them, denoted as $\mathbbm{L}$.
When they further collect information from $v$ semi-honest users, the logins initiated by these users are picked out of $\mathbbm{L}$ and linked together as
$\mathfrak{L}^m=\left \{ \begin{matrix}
L^m_{1,1},&L^m_{1,2},&\cdots,&L^m_{1,c}\\
L^m_{2,1},& L^m_{2,2},&\cdots,&L^m_{2,c}\\
\cdots,&\cdots,&L^m_{i,j},&\cdots\\
L^m_{v,1},&L^m_{v,2},&\cdots,&L^m_{v,c}
\end{matrix}\right\}$,
where $L^m_{i, j}=([r_j]G, t_{i,j}, [u_ir_j]G) \in \mathbbm{L}$ for $1 \le i \le v$ and $1 \le j \le c$. Any login in $\mathbbm{L}$ but not linked in $\mathfrak{L}^m$ is initiated by an honest user to visit one of the $c$ semi-honest RPs.

\begin{figure}[tb]
  \centering
  \includegraphics[width=1.0\linewidth]{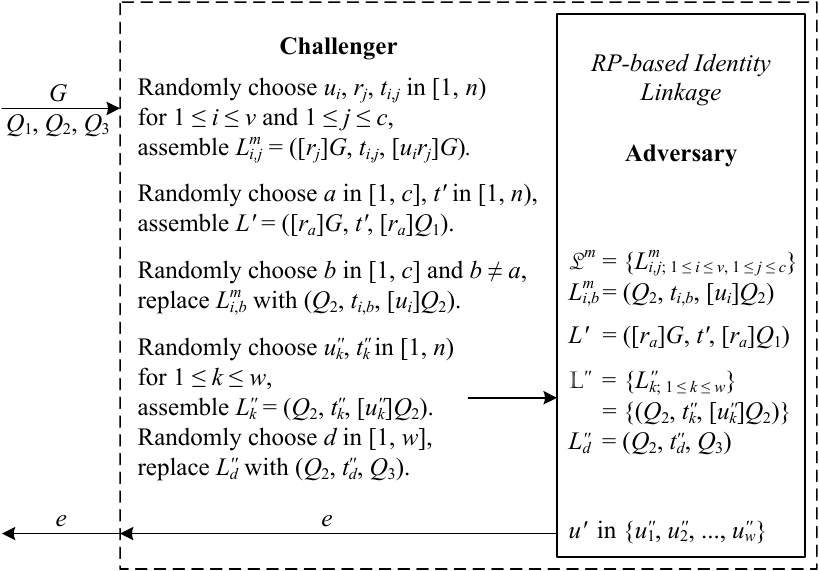}
  \caption{The PPT algorithm $\mathcal{D}^*_R$ constructed based on the RP-based identity linkage game to solve the ECDDH problem.}
  \label{fig:dalgorithm}
\end{figure}

In Theorem \ref{thm-rp-unlinkability} we prove that, 
even when a number of semi-honest users share information with them,
\emph{the semi-honest RPs still cannot link any login visiting a semi-honest RP by an honest user to any other logins visiting other semi-honest RPs by honest users}.
Then, RP unlinkability is ensured.

\vspace{1.5mm}
\begin{thm}[RP Unlinkability]
\emph{Given $\mathbbm{L}$ and $\mathfrak{L}^m$, $c$ semi-honest RPs and $v$ semi-honest users cannot link any login visiting some of these RPs by an honest user to any subset of logins visiting other RPs by honest users.}
\label{thm-rp-unlinkability}
\end{thm}

\noindent\textbf{\textsc{Proof.}} Out of $\mathbbm{L}$
 we randomly choose a login $L' \neq L^m_{i,j}$ $(1 \le i \le v, 1 \le j \le c)$,
 which is initiated by an (unknown) honest user with $ID_{U'}=u'$ to a semi-honest $RP_{a}$ where $a \in [1,c]$.
Then, we randomly choose another semi-honest $RP_{b}$, where $b \in [1,c]$ and $b \neq a$.
Consider any subset $\mathbbm{L}'' \subset \mathbbm{L}$ of $w$ $(1 \leq w < s-v)$ logins visiting $RP_{b}$ by unknown honest users,
 and denote the identities of the users initiating these logins as $\mathbbm{u}_w=\{{u''_1}, {u''_2}, \cdots, {u''_w}\} \subset \mathbbm{u}$.

Next, we prove that the group of semi-honest adversaries cannot distinguish $u' \in \mathbbm{u}_w$ from $u'$ randomly selected in the universal set (i.e., $u' \in \mathbb{Z}_n$).
So the adversaries cannot link $L'$ to another login (or any subset of logins) visiting $RP_{b}$.

We define an RP-based identity linkage game $\mathcal{G}_R$ between an adversary and a challenger, to describe this login linkage: the adversary receives $\mathfrak{L}^m$, $L'$, and $\mathbbm{L}''$ from the challenger and outputs $e$, where (\emph{a}) $e = 1$ if it decides $u'$ is in $\mathbbm{u}_w$ 
or (\emph{b}) $e=0$ if it believes $u'$ is randomly chosen from $\mathbb{Z}_n$.
Thus, the adversary succeeds in $\mathcal{G}_R$ with an advantage $\mathbf{Adv}$:
\begin{align*}
&{\rm Pr}_1={\rm Pr}(\mathcal{G}_R(\mathfrak{L}^m, L', \mathbbm{L}'')=1 \;| \; u' \in \mathbbm{u}_w)  \\
&{\rm Pr}_2={\rm Pr}(\mathcal{G}_R(\mathfrak{L}^m, L', \mathbbm{L}'')=1 \; | \; u' \in \mathbb{Z}_n)\\
&{\mathbf{Adv}}=|{\rm Pr}_1-{\rm Pr}_2|
\end{align*}

As depicted in Figure \ref{fig:dalgorithm}, we design a PPT algorithm $\mathcal{D}^*_R$ based on $\mathcal{G}_R$ to solve the elliptic curve decisional Diffie-Hellman (ECDDH) problem: Given $(G, [x]G$, $[y]G$, $[z]G)$, decide whether $z$ is equal to $xy$ or randomly chosen in $\mathbb{Z}_n$, where $G$ is a point on an elliptic curve $\mathbb{E}$ of order $n$, and $x$ and $y$ are integers randomly and independently chosen in $\mathbb{Z}_n$.

The algorithm $\mathcal{D}^*_R$ works as follows. (1) On receiving an input $(G, Q_1=[x]G, Q_2=[y]G, Q_3=[z]G)$,
the challenger
chooses random numbers in $\mathbb{Z}_n$ to construct $\{u_i\}$, $\{r_j\}$, and $\{t_{i, j}\}$ for $1 \le i \le v$ and $1 \le j \le c$, with which it assembles $L^m_{i, j}=([r_j]G, t_{i,j}, [u_ir_j]G)$.
It ensures $[r_{j}]G \neq Q_2$ (or $r_j \neq y$) in this procedure.
(2) It randomly chooses $a \in [1, c]$ and $t' \in \mathbb{Z}_n$, to assemble $L' = ([r_{a}]G, t', [r_{a}]Q_1) = ([r_{a}]G, t', [xr_{a}]G)$.
(3)
Next, the challenger randomly chooses $b \in [1, c]$ but $b \neq a$, and replaces $ID_{RP_b}$ with $Q_2 = [y]G$.
Hence, for $1 \le i \le v$, the challenger replaces $L^m_{i, b}=([r_b]G, t_{i,b}, [u_ir_b]G)$ with $(Q_2, t_{i,b}, [u_i]Q_2) = ([y]G, t_{i,b}, [u_iy]G)$, and finally constructs $\mathfrak{L}^m$.
(4) The challenger chooses random numbers in $\mathbb{Z}_n$ to construct $\{u''_k\}$ and $\{t''_k\}$ for $1 \leq k \leq w$,
 with which it assembles $\mathbbm{L}'' = \{L''_{k; 1\leq k \leq w}\} = \{(Q_2, t''_k, [u''_k]Q_2)\} = \{([y]G, t''_k, [u''_ky]G)\}$.
It ensures $[u''_k]G \neq Q_1$ (i.e., $u''_k \neq x$) and $u''_k \neq u_i$,
 for $1 \le i \le v$ and $1 \le k \le w$.
Finally, it randomly chooses $d \in [1, w]$ and replaces $L''_{d}$ with $(Q_2, t''_d, Q_3) = ([y]G, t''_d, [z]G)$.
 Thus, $\mathbbm{L}'' = \{L''_{k;1\leq k \leq w}\}$ represents the logins initiated by $w$ honest users, i.e., $\mathbbm{u}_w=\{u''_1, u''_2, \cdots, u''_{d-1}, z/y, u''_{d+1}, \cdots, u''_w\}$.
 (5) When the adversary of $\mathcal{G}_R$ receives $\mathfrak{L}^m$, $L'$, and $\mathbbm{L}''$ from the challenger, it returns $e$ which is also the output of $\mathcal{D}^*_R$.

According to the above construction, 
$x$ is embedded as $ID_{U'}$ of the login $L'$ visiting the RP with $ID_{RP_{a}} = [r_{a}]G$,
and $z/y$ is embedded as $ID_{U''_d}$ of $\mathbbm{L}''$ visiting the RP with $ID_{RP_{b}}=[y]G$,
together with $\{u''_1, \cdots, u''_{d-1}, u''_{d+1}, \cdots, u''_w\}$.
Meanwhile, $[r_{a}]G$ and $[y]G$ are two semi-honest RPs' identities in $\mathfrak{L}^m$.
Because $x \neq u''_{k; 1\leq k \leq w, k \neq d}$ and then $x$ is not in $\{u''_1, \cdots, u''_{d-1}, u''_{d+1}, \cdots, u''_w\}$, the adversary outputs $s=1$ and succeeds in this game only if $x = z/y$.
Therefore, using $\mathcal{D}^*_R$ to solve the ECDDH problem, we have an advantage $\mathbf{Adv}^*=|{\rm Pr}^*_1 - {\rm Pr}^*_2|$, where
\begin{align*}
&{\rm Pr}^*_1 =  {\rm Pr}(\mathcal{D}^*_R(G, [x]G, [y]G, [xy]G)=1) \\
=&{\rm Pr}(\mathcal{G}_R(\mathfrak{L}^m, L', \mathbbm{L}'')=1 \; | \; u' \in \mathbbm{u}_w) = {\rm Pr}_1 \\
&{\rm Pr}^*_2= {\rm Pr}(\mathcal{D}^*_R(G, [x]G, [y]G, [z]G)=1) \\
=&{\rm Pr}(\mathcal{G}_R(\mathfrak{L}^m, L', \mathbbm{L}'')=1 \; | \; u' \in \mathbb{Z}_n) = {\rm Pr}_2 \\
&\mathbf{Adv}^*=|{\rm Pr}^*_1-{\rm Pr}^*_2|=|{\rm Pr}_1-{\rm Pr}_2|={\mathbf{Adv}}
\end{align*}

If the adversary of $\mathcal{G}_R$ has a non-negligible advantage, then $\mathbf{Adv}^*={\mathbf{Adv}}$ is also non-negligible regardless of the security parameter $\lambda$. This violates the ECDDH assumption.

So the adversary has no advantage in $\mathcal{G}_R$ and cannot decide whether $L'$ is initiated by some honest user with an identity in $\mathbbm{u}_w$ or not.
Because $RP_b$ is any semi-honest RP, this proof can be extended from $RP_b$ to more semi-honest RPs.
\hfill $\square$

\subsection{Account Synchronization}
\label{account-syn}
If new users are allowed to register after the initialization,
to distinguish meaningless accounts from meaningful ones in Step 3.2,
 an RP needs to synchronize the accounts from the IdP to maintain an \emph{up-to-date} set of meaningful accounts;
otherwise, 
  a \emph{newly-registered} user will be rejected
    because the derived account is \emph{falsely} considered as meaningless (i.e., not in its local set).

Naive account synchronization may be finished as below.
On receiving an identity token from which a meaningless account is derived,
    an RP instantly contacts the IdP to retrieve the account set of users registered after the RP's last synchronization.
Then, after this instant account synchronization,\footnote{An RP does not straightforwardly treat the token holder with an account not in its local set
 as a newly-registered user;
otherwise,
although a malicious user cannot log into this RP as any meaningful account belonging to others according to Theorem \ref{thm-u-id},
 she could log into such an RP as an meaningless but identical account in multiple logins.
For example,
    by choosing ($\hat{t}_1$, $\check{t}_1$) and ($\hat{t}_2$, $\check{t}_2$) in two logins
     where $m=\hat{t}_1/\check{t}_1 = \hat{t}_2/\check{t}_2$,
    as analyzed in the proof of Theorem \ref{thm-u-id}.
this malicious user would actually own another account $Acct=[mID_U]ID_{RP} \neq [ID_U]ID_{RP}$.} 
    the token holder will be rejected
     if the derived account is still not in the updated set.


The following ``silient'' account synchronization
     leverages \emph{conditional} $PID_{RP}$ checking as below,
eliminating the communications in the ``active'' synchronization as above.
When an RP derives some account not in its local account set,
    it additionally checks $PID_{RP}$ in this case:
     If $PID_{RP}$ enclosed in the signed token is equal to $[t]ID_{RP}$,
                this account is asserted to belong to a newly-registered user (i.e., meaningful)
                and then the RP locally updates its account set (i.e., this derived account is inserted into the set);
    otherwise, it rejects this token binding $PID_{RP} \neq [t]ID_{RP}$.
In this case, adversaries cannot (\emph{a}) manipulate $t$ to visit the designated RP because $t$ is checked in the equation of $PID_{RP} = [t]ID_{RP}$ 
 or (\emph{b}) induce any RP other than the designated one to accept this token
    because they cannot solve $t'$ satisfying that $[t']ID_{RP'} = [t]ID_{RP}$, which is equivalent to solve the ECDLP.
So the derived account must be meaningful if $PID_{RP}$ is equal to $[t]ID_{RP}$.

\section{Implementation and Evaluation}
\label{sec:implementation}

We implemented the UPPRESSO prototype\footnote{The prototype is open-sourced at \url{https://github.com/uppresso/}.} and conducted experimental comparisons with two open-source SSO systems:
 (\emph{a}) MITREid Connect \cite{MITREid}, a PPID-enhanced OIDC system with redirect UX,
  preventing only RP-based identity linkage,
 and (\emph{b}) SPRESSO \cite{SPRESSO}, which prevents only IdP-based login tracing.
All these solutions work with COTS browsers as user agents,
    while UPPRESSO and SPRESSO implement OIDC-compatible services with pop-up UX.

\subsection{Prototype Implementation}
\label{subsec:proto-imple}

The UPPRESSO prototype implemented the identity transformations on the NIST P256 elliptic curve where $n \approx 2^{256}$,
and the IdP was developed on top of MITREid Connect \cite{MITREid}, 
with minimal code modifications.
All systems employ RSA-2048 and SHA-256 to generate tokens.
The scripts of user-i and user-r consist of about 180 and 80 lines of JavaScript code, respectively.  
The cryptographic computations such as $Cert_{RP}$ verification and $PID_{RP}$ negotiation are conducted based on elliptic \cite{elliptic-lib}, an open-source JavaScript library.

We developed a Java-based RP SDK with about 310 lines of code on the Spring Boot framework.
The main function encapsulates the RP operations of UPPRESSO: verify an identity token and derive $Acct$. The cryptographic computations are finished using the Spring Security library.
Then, an RP can invoke necessary functions by adding less than 10 lines of Java code, to access the services provided by UPPRESSO.

SPRESSO implements all entities by JavaScript based on node.js, while MITREid Connect provides Java implementations of IdP and RP SDK.
Thus, in UPPRESSO and MITREid Connect, we implemented RPs based on Spring Boot by integrating the respective SDKs. In all three schemes, the RPs provide the same function of obtaining the user's account from a verified identity token.


\subsection{Performance Evaluation}
\label{sec:evaluation}

\begin{figure*}[tb]  
  \centering
    \renewcommand{\thesubfigure}{[\roman{subfigure}]} 
  \subfigure[\footnotesize Initial visit]{
    \begin{minipage}[b]{0.85\columnwidth}
      \includegraphics[width=\linewidth]{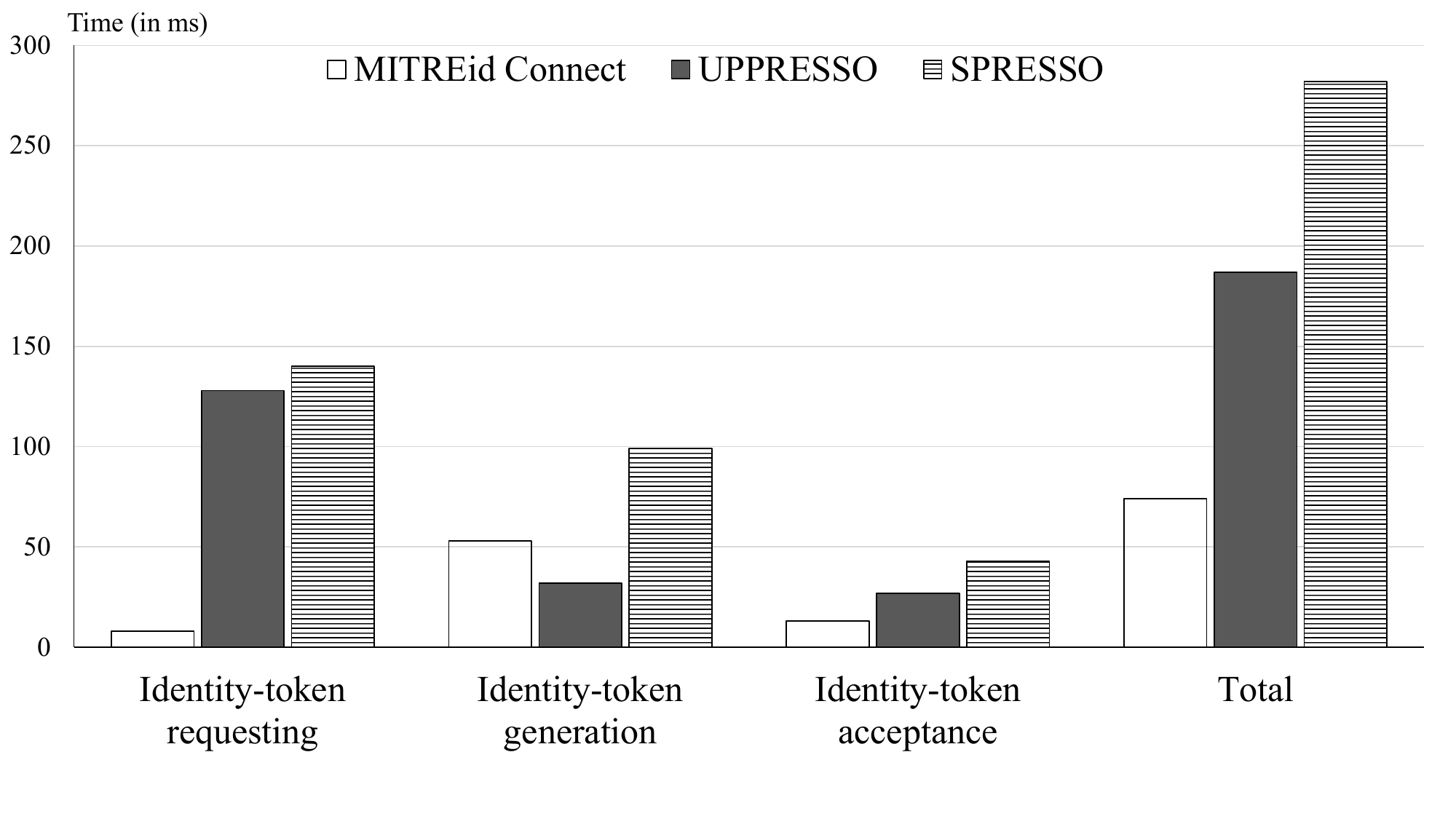}
    \end{minipage}
  }\hfill
  \subfigure[\footnotesize Subsequent visit]{
    \begin{minipage}[b]{0.85\columnwidth}
      \includegraphics[width=\linewidth]{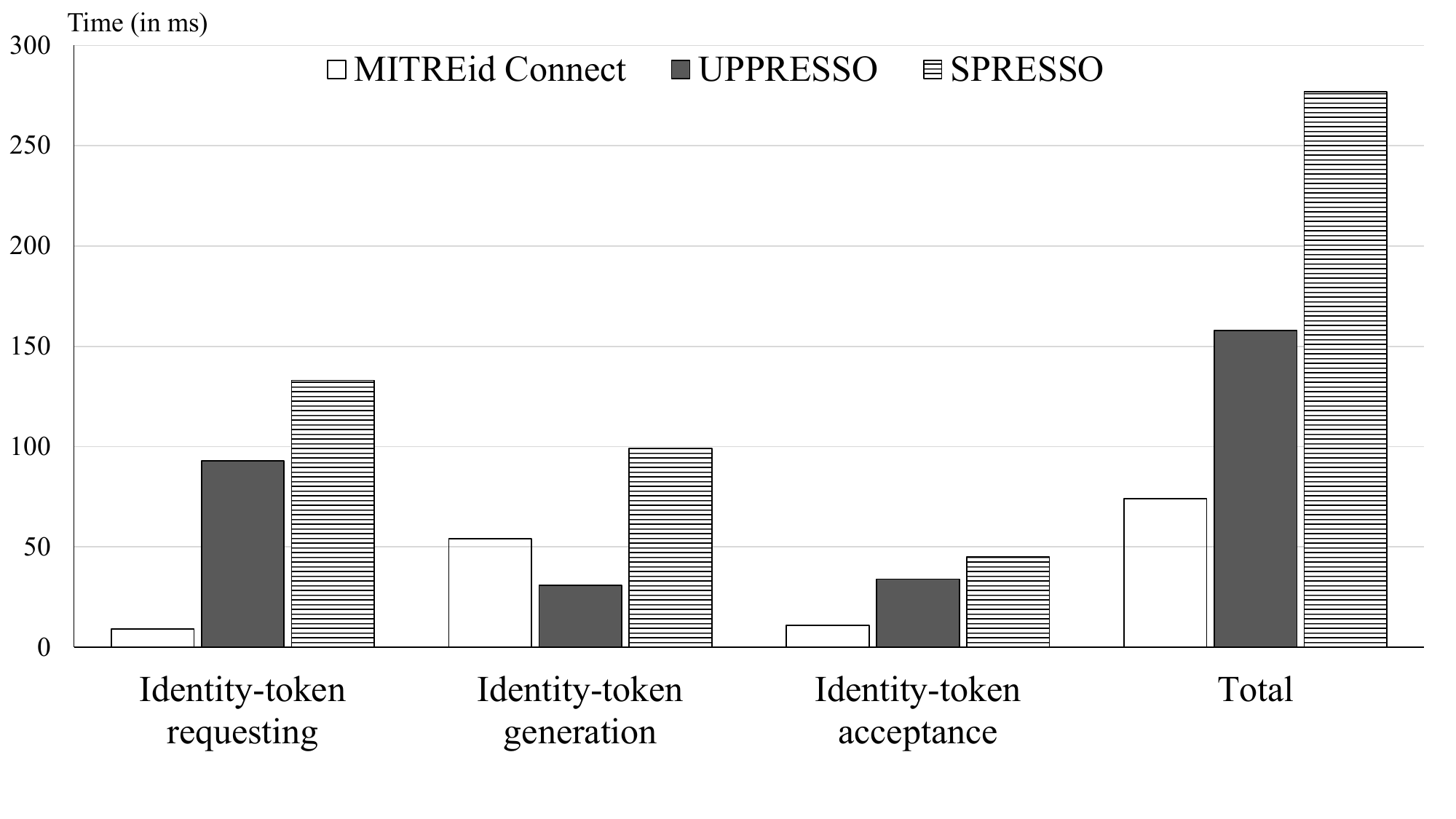}
    \end{minipage}
  }
  \vspace*{-6pt}
  \par\medskip
  \centering{\footnotesize (\emph{a}) Within a local virtual environment}

    \setcounter{subfigure}{0} 
  \subfigure[\footnotesize Initial visit]{
    \begin{minipage}[b]{0.85\columnwidth}
      \includegraphics[width=\linewidth]{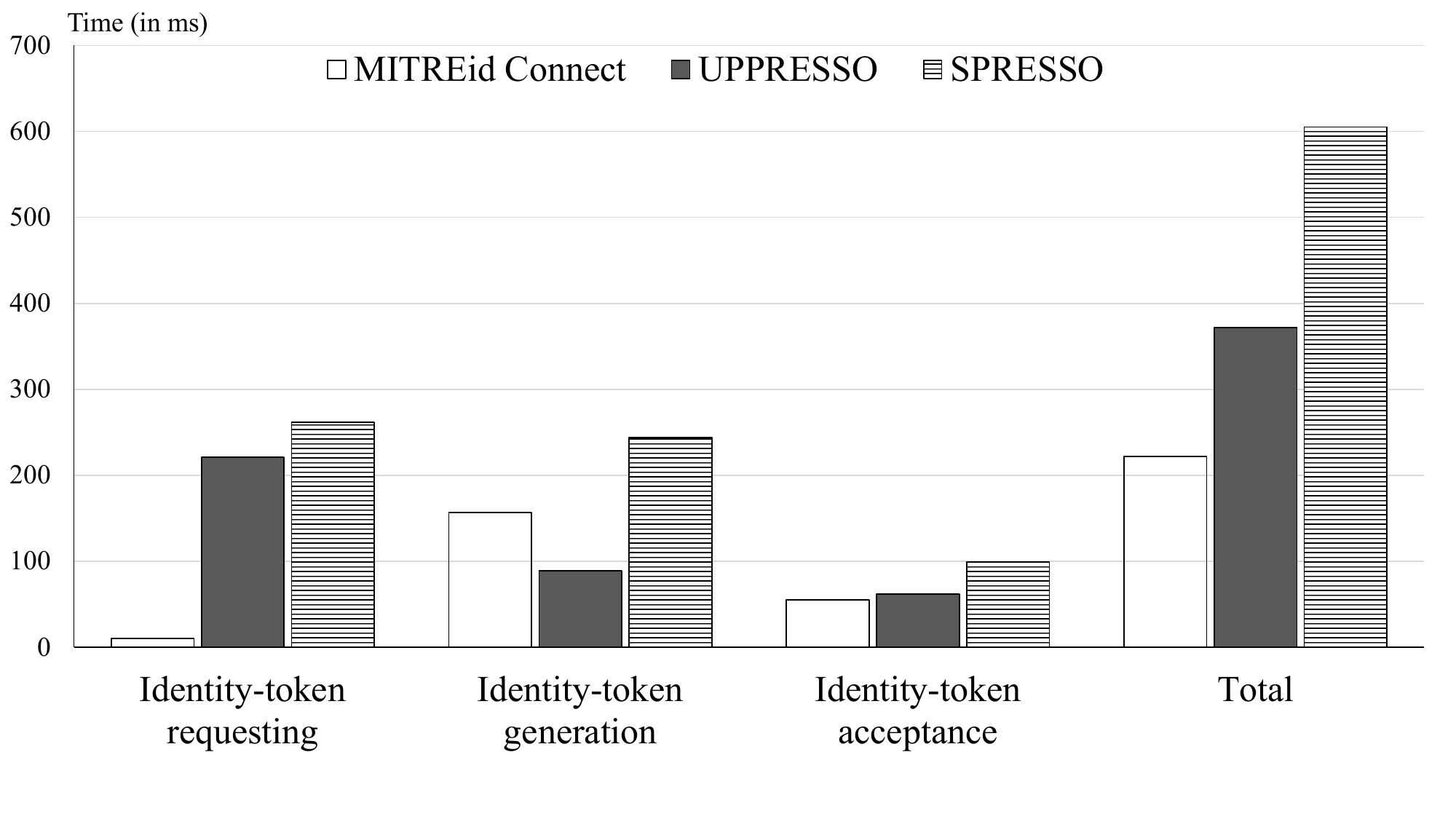}
    \end{minipage}
  }\hfill
  \subfigure[\footnotesize Subsequent visit]{
    \begin{minipage}[b]{0.85\columnwidth}
      \includegraphics[width=\linewidth]{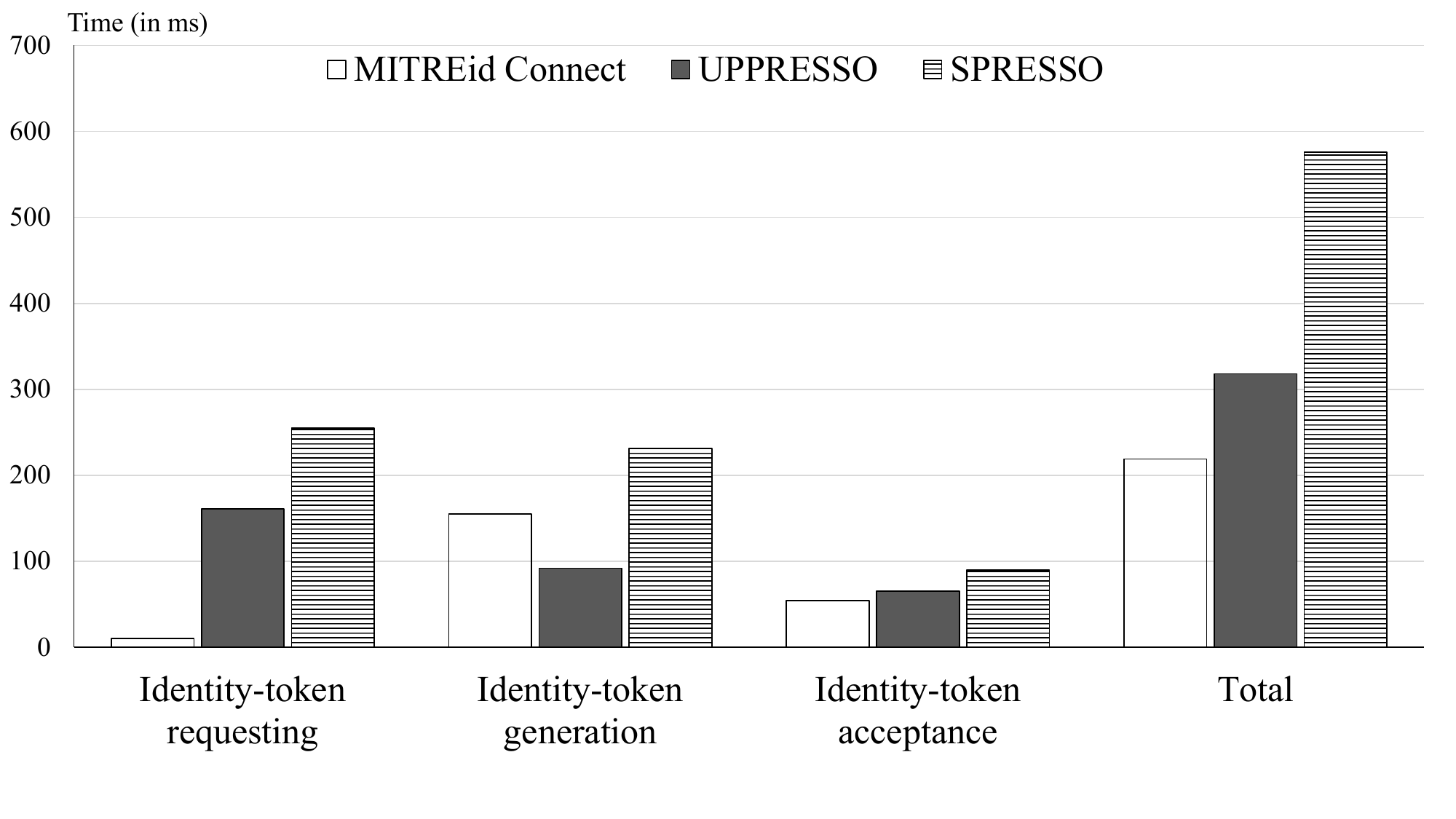}
    \end{minipage}
  }
  \vspace*{-6pt}
  \par\medskip
  \centering{\footnotesize (\emph{b}) With a remotely-visiting browser}

  \caption{The time costs of SSO login in MITREid Connect, UPPRESSO, and SPRESSO}
  \label{fig:evaluation}
\end{figure*} 

\noindent {\bf Experiment setting.}
We conducted experiments in two settings: (\emph{a}) on a PC with AMD Ryzen 7 57000X CPU and 32GB memory, a browser, the IdP, the visited RP, and the extra forwarder server of SPRESSO were deployed on separated virtual machines, each of which ran Windows Server 2025 with 2 vCPUs and 4GB RAM, and (\emph{b}) the browser running locally on the PC with AMD Ryzen 7 57000X CPU and 32GB memory, remotely accessed the servers (IdP and RP servers, and the forwarder server of SPRESSO) deployed on the Alibaba Cloud Elastic Compute Service, each of which ran Windows Server 2025 with 8 vCPUs and 32GB RAM. 
The first setting ran all functions on separated virtual machines but within the same physical PC,
    to minimize the impact of network delays.



\noindent {\bf Comparisons.} We split the login flow into three phases for detailed comparisons: (1)
{\em identity-token requesting} (Step 1 in Figure \ref{fig:process}), to construct an identity-token request and send it to the IdP server; (2) {\em identity-token generation} (Step 2), to generate an identity token at the IdP server, while the user authentication and the user-attribute consent are excluded; and (3) {\em identity-token acceptance} (Step 3), where the RP receives, verifies, and parses the identity token.


The IdP and the RP of all solutions, as well as the forwarder server in SPRESSO,
     require user-side scripts, which introduces additional time for downloading. 
In the identity-token requesting phase of UPPRESSO a browser downloads the user-i script,
    as described in Section \ref{sec:web-design}.
As mentioned in Section \ref{subsec:OIDC},
 to process the token retrieved from the IdP, 
    in MITREid Connect a user-r script is downloaded during the phase of token generation.
In addition to a script from the IdP in identity-token requesting,
SPRESSO needs another script from the forwarder server in the token acceptance phase \cite{SPRESSO}.
Fortunately, once a browser visits a web server, it caches the scripts locally, reducing the download time for subsequent visits.
This caching is enabled by the \verb+localStorage+ HTML object \cite{htmlStorageHtml} and HTTP  \texttt{Cache-Control} header.

We measured the time cost for both initial and subsequent visits (1,000 trials) and compared the average SSO login times of three schemes.
As shown in Figure \ref{fig:evaluation},
MITREid Connect, UPPRESSO, and SPRESSO require (\emph{a}) 74 ms, 187 ms, and 282 ms for initial visits, and 69 ms, 158 ms and 277 ms for subsequent visits,
 or (\emph{b}) 222 ms, 372 ms, and 605 ms for initial visits, and 219 ms, 318 ms and 576 ms for subsequent visits.

Regarding identity-token requesting, 
the RP of MITREid Connect immediately constructs an identity-token request. 
UPPRESSO incurs overheads in opening a new browser window and downloading the user-i script.
%
%
In SPRESSO the user also needs to open a new browser window, and obtain information about the IdP 
and encrypt the visited RP's domain using an ephemeral key, resulting in extra overheads.

UPPRESSO requires the least time for generating identity tokens for it receives the token from the IdP without additional processing.
In this phase MITREid Connect and SPRESSO require extra time as the browser downloads a script from the RP and the forwarder, respectively. 
After receiving an identity token from the IdP, in SPRESSO the user decrypts the RP's domain by scripts and sends it to the RP's redirection endpoint.
Moreover, SPRESSO takes slightly more time to generate an identity token, as it implements the IdP using node.js and uses a JavaScript cryptographic library that is a little less efficient than the Java library used in the others.

Finally, in the identity-token acceptance phase, 
three solutions take similar amounts of time for the RP to receive a token and accept it.

By comparing the average time costs of initial visits and subsequent visits,
    we find that the performances of all schemes are improved
        and the improvement is the most significant in UPPRESSO for the size of its scripts is the largest (134kB vs. 2.34kB in MITREid Connect and 6.84kB in SPRESSO). 
\section{Discussions}
\label{sec:discussion}

\noindent \textbf{IdP-RP collusive attacks.} 
This work prefers to provide better user convenience by easy-to-deploy privacy-preserving designs,
rather than prevent IdP-RP collusive attacks.
Unless a user secret is involved in the derivation of accounts,
SSO cannot protect user privacy against an IdP colluding with RPs.
For example, even when the IdP strictly follows the protocols but shares information with RPs,
    a user would complete her logins \emph{entirely} with colluding entities,
    and then the IdP and RPs could always link a user's (pseudo-)identities and accounts (i.e., both IdP untraceability and RP unlinkability are broken).
That is, regardless of which $\mathcal{F}_{Acct\ast}()$ is adopted to determine
$Acct_{i,j} = \mathcal{F}_{Acct\ast}(ID_{U_i}, ID_{RP_j})$,
the IdP could always collude with RPs to track a user's login activities by analyzing the relationship between the user's identity and accounts derived at the RPs,
unless $Acct_{i,j} = \mathcal{F}_{Acct\ast}(s_i,ID_{U_i}, ID_{RP_j})$ is adopted and $s_i$ is a long-term user secret unknown to the colluding IdP and RPs.

As mentioned in Section \ref{subsec-solutions}, privacy-preserving identity federation \cite{ELPASSO, UnlimitID, idemix, PseudoID, uprov} prefers stronger privacy than user convenience,
    where a user secret is introduced to mask the relationship among the (pseudo-)identities and accounts.
In this case,
    a user needs to install a browser extension 
            to maintain the secret.
Moreover, if the user accesses the services from different devices,
    additional operations are required to synchronize the secret among these devices.

Alternatively, MISO \cite{miso} introduces an extra fully-trusted mixer server other than the IdP to keep $s_i$,
            to calculate user pseudo-identities (or accounts) in each login.
Then, in MISO the mixer server can track all users' login activities,
   either colluding with RPs or not.





\noindent \textbf{Support for the authorization code flow.} In the authorization code flow of OIDC \cite{OpenIDConnect}, the IdP does not directly return identity tokens.
Instead, it generates an authorization code, which is forwarded to the target RP.
The RP uses this code to retrieve identity tokens from the IdP.

$\mathcal{F}_{Acct\ast}()$, $\mathcal{F}_{PID_{U}}()$, $\mathcal{F}_{PID_{RP}}()$ and $\mathcal{F}_{Acct}()$, can be also integrated into the authorization code flow of OIDC to transform (pseudo-)identities in signed identity tokens.
Then, in the flow an authorization code (but not the token) will be sent to the user-r script, and to the RP finally.
This code serves as an index to retrieve tokens by the RP, not disclosing any information about the user.
On receiving an authorization code, the RP uses it as well as a credential issued by the IdP during the initial registration \cite{OpenIDConnect}, to retrieve identity tokens from the IdP's token endpoint \cite{rfc6749}.

Moreover, to hide RP identities from the IdP in the retrieval of identity tokens in this flow,
 privacy-preserving credentials
    (e.g., ring signatures \cite{ring-sig} or privacy passes \cite{privacypass,trusttoken})
 and anonymous communications (e.g., Tor \cite{tor} or oblivious proxies \cite{ohttp-rfc}) need to be adopted for RPs.

\noindent \textbf{Support for redirect UX.}
With redirect UX, 
    an RP and the IdP exchange messages via HTTP redirection within one browser window.
If we implement the UPPRESSO protocol with redirect UX,
the following designs are needed to hide the visited RP's identity from the IdP 
during these HTTP redirections.
These designs work either in the implicit flow of OIDC or the authorization code flow.

In an OIDC system without the protections of user privacy,
the RP's identity and redirection endpoint URL (i.e., $ID_{RP}$ and $ReEnptUrl_{RP}$) are sent as query parameters in the HTTP redirection to the IdP, 
    to request identity tokens.
On the contrary, in UPPRESSO
 when a user clicks and redirects the browser from the RP-login page to the IdP server,
    $ID_{RP}$ and $ReEnptUrl_{RP}$ are not sent.
Meanwhile,
    $Cert_{RP}$ and the scope of requested user attributes are carried as URL fragments,
so they are processed locally within the browser and not sent to the IdP web server.
Besides, the RP suppresses the HTTP \texttt{referer} header (e.g., by setting \verb+rel="noreferrer"+) of this redirection, which prevents the IdP server from receiving the RP's domain in the HTTP request of this redirection.

In the redirected page
the user-i script verifies $Cert_{RP}$, extracts $ID_{RP}$ and $ReEnptUrl_{RP}$,
calculates $PID_{RP}$, and then requests an identity token from the IdP's authorization endpoint.
Then, to forward the signed token to the RP,
    in the original OIDC protocol
        the IdP server constructs a server-side HTTP redirection to $ReEnptUrl_{RP}$.
However, in UPPRESSO the IdP does not know the visited RP,
    so it returns the signed token to the user-i script but with an empty $ReEnptUrl_{RP}$.
Then, the user-i script loads $ReEnptUrl_{RP}$ and $t$ locally from \verb+sessionStorage+, and submits the identity token and also $t$ to the RP's redirection endpoint 
by a client-side redirection (e.g., by \verb+window.location.href+).

%

\noindent \textbf{Alternatives to generate $\boldsymbol{ID_{RP}}$ and bind $\boldsymbol{ReEnptUrl_{RP}}$.}
In UPPRESSO the IdP generates random $ID_{RP}$ and signs an RP certificate to bind $ID_{RP}$ and $ReEnptUrl_{RP}$, which is verified by the user-i script
    and then ensures the relationship between the RP designated by an identity token and the URL to receive this token.
It also guarantees that the target RP has already registered itself at the IdP and prevents unregistered RPs without such certificates from utilizing the IdP's services.

An alternative method to bind $ID_{RP}$ and $ReEnptUrl_{RP}$ is
 a deterministic scheme to calculate unique $ID_{RP}$ based on the RP's unambiguous name such as its domain.
This can be achieved by encoding the domain with a hashing-to-elliptic-curves function \cite{irtf-cfrg-hash-to-curve-16}, which provides collision resistance but not revealing the elliptic-curve discrete logarithm of the output. It generates a point on $\mathbb{E}$ as $ID_{RP}$, ensuring the \emph{uniqueness} of $ID_{RP} = [r]G$ while keeping $r$ unknown. 
In this case, in Step 1.2 the user-r script sends only the RP's redirection endpoint URL but not its RP certificate, and the user-i script calculates $ID_{RP}$ by itself based on the RP's domain. 
Then, any RP can be visited through the SSO services, either registered or not.

However, if some RP changes its domain, for example, from \verb+https://theRP.com+ to \verb+https://RP.com+, the accounts (i.e., $Acct = [ID_U]ID_{RP}$) will inevitably change.
Thus, a user is required to perform special operations to migrate her account to the updated RP system.
It is worth noting that the user operations cannot be eliminated in the migration to the updated one;
otherwise, it implies two colluding RPs could link a user's accounts across these RPs.

\noindent \textbf{Restriction of the targetOrigin mechanism.}
The user-i script forwards tokens to the user-r script, and the \verb+postMessage+ targetOrigin mechanism \cite{postm-targeto} restricts the recipient of the forwarded identity tokens, to ensure that the tokens will be sent to the intended $ReEnptUrl_{RP}$, as specified in the RP certificate. The targetOrigin is specified as a combination of protocol, port (if not present, 80 for \verb+http+ and 443 for \verb+https+), and domain (e.g., \verb+RP.com+).
The origin of the user-r script must accurately match the targetOrigin to receive tokens.

The targetOrigin mechanism does not check the whole URL path in $ReEnptUrl_{RP}$, but it introduces no additional risk.
Consider two RPs in one domain but receiving tokens through different URLs,
 e.g., \verb+https://RP.com/honest/tk+ and \verb+https://RP.com/malicious/tk+.
This mechanism cannot distinguish them.
%
%
Because a browser controls access to web resources following the same-origin policy (SOP) \cite{sop},
    a user's resources in the honest RP web server is accessible to  the malicious server.
 For example, it could steal cookies by \verb+window.open('https://RP.com/honest').+ \verb+document.cookie+,
 even if the honest RP restricts only the HTTP requests to specific paths are allowed to access its cookies.
So this risk is caused by the SOP model of browsers but not our designs,
    and commonly exists in SSO services accessed from COTS browsers \cite{rfc6749,SPRESSO,MITREid,GoogleIdIntegrate,uber}.



\section{Conclusion}
\label{sec:conclusion}
This paper presents UPPRESSO, an untraceable and unlinkable privacy-preserving SSO system for protecting a user's online profile across RPs against both a curious IdP and colluding RPs.
We propose an identity-transformation approach and design algorithms satisfying the requirements of security and privacy: (\emph{a}) $\mathcal{F}_{Acct\ast}()$ determines a user's account, unique at an RP and unlinkable across RPs,
(\emph{b}) $\mathcal{F}_{PID_{RP}}()$ protects a visited RP's identity from the curious IdP, (\emph{c}) $\mathcal{F}_{PID_{U}}()$ prevents colluding RPs from linking a user's multiple logins visiting different RPs, and (\emph{d}) $\mathcal{F}_{Acct}()$ derives a user's permanent account from ephemeral pseudo-identities.

The identity transformations are integrated into the widely-adopted OIDC protocol, while maintaining user convenience and security guarantees of SSO services. Our experimental evaluations of the UPPRESSO prototype demonstrate its efficiency.
For example,
            when the scripts are not cached,
in UPPRESSO a login takes 187 ms in average when the IdP, the visited RP, and user browsers are deployed in a virtual private cloud, or 372 ms when a user visits remotely.
Compared with MITREid Connect,
it introduces an overhead of 89 -- 150 ms to prevent two privacy threats.







%


\bibliographystyle{IEEEtran}
\bibliography{ref}

@misc{understanding-arxiv,
    author = {J. Lin and B. Zhang and W. Wang and Q. Cai and J. Jing and H. He},
    title = {{Understanding the Identity-Transformation Approach in OIDC-Compatible Privacy-Preserving SSO Services}},
    howpublished = {\url{https://arxiv.org/abs/2506.01325}},
    year = {2025}}

@manual{url-fragment,
  author    = {T. Berners-Lee and R. Fielding and L. Masinter},
  title     = {{RFC 3986: Uniform resource identifier (URI): Generic syntax}},
  organization={Internet Engineering Task Force},
  year      = {2005},
}

@inproceedings{up-sso,
    title = {{UP-SSO: Enhancing the user privacy of SSO by integrating PPID and SGX}},
    author = {C. Guo and F. Lang and Q. Wang and J. Lin},
    booktitle = {International Conference on Advanced Computing and Endogenous Security (ICACES)},
    year = {2021}}

@inproceedings{sok-oprf,
    author = {S. Casacuberta and J. Hesse and A. Lehmann},
    title = {{SoK: Oblivious pseudorandom functions}},
    booktitle = {7th IEEE European Symposium on Security and Privacy (EuroS\&P)},
    year = {2022}}

@misc{uppresso-arxiv,
    author = {C. Guo and J. Lin and Q. Cai and W. Wang and F. Li and Q. Wang and J. Jing and B. Zhao},
    title = {{UPPRESSO: Untraceable and unlinkable privacy-preserving single sign-on services}},
    howpublished = {\url{https://arxiv.org/abs/2110.10396}},
    year = {2022}}

@inproceedings{ZKP-GPU,
    title = {{GZKP: A GPU accelerated zero-knowledge proof system}},
    author = {W. Ma and Q. Xiong and X. Shi and X. Ma and H. Jin and H. Kuang and M. Gao and Y. Zhang and H. Shen and W. Hu},
    booktitle = {28th ACM International Conference on Architectural Support for Programming Languages and Operating Systems (ASPLOS)},
    pages = {340-353},
    year = {2023}}

@inproceedings{POIDC,
    title = {{Privacy-preserving OpenID Connect}},
    author = {S. Hammann and R. Sasse and D. Basin},
    booktitle = {15th ACM Asia Conference on Computer and Communications Security (AsiaCCS)},
    pages={277-289},
    year = {2020}}

@article{save-flow,
  author    = {M. Kroschewski and A. Lehmann},
  title     = {{Save the implicit flow? Enabling privacy-preserving RP authentication in OpenID Connect}},
  journal   = {Privacy Enhancing Technologies},
  volume    = {2023},
  number    = {4},
  pages     = {96-116},
  year      = {2023},
}

@article{TSAPP,
    author ={Z. Zhang and C. Xu and C. Jiang and K. Chen},
    title = {{TSAPP: Threshold single-sign-on authentication preserving privacy}},
    journal   = {IEEE Transactions on Dependable and Secure Computing},
    volume = {21},
    number={4},
    pages={1515-1527},
    year = {2024}}

@inproceedings{miso,
  author    = {R. Xu and S. Yang and F. Zhang and Z. Fang},
  title     = {{MISO: Legacy-compatible privacy-preserving single sign-on using trusted execution environments}},
  booktitle = {8th IEEE European Symposium on Security and Privacy (EuroS\&P)},
  year      = {2023}}

@inproceedings{oprf-deduplication,
    title = {{Oblivious PRF on committed vector inputs and application to deduplication of encrypted data}},
    author = {J. Camenisch and A. de Caro and E. Ghosh and A. Sorniotti},
    booktitle = {23rd International Conference on Financial Cryptography and Data Security (FC)},
    year = {2019},
    pages = {337-356}}

@inproceedings{o-kms,
    title = {{Updatable oblivious key management for storage systems}},
    author = {S. Jarecki and H. Krawczyk and J. Resch},
    booktitle = {26th ACM Conference on Computer and Communications Security (CCS)},
    year = {2019},
    pages = {379-393}}

@inproceedings{Private-Contact-Discovery,
    title = {Mobile private contact discovery at scale},
    author = {D. Kales and C. Rechberger and T. Schneider and M. Senker and C. Weinert},
    booktitle = {28th USENIX Security Symposium},
    year = {2019}}

@inproceedings{pp-ss,
    title = {{Password-protected secret sharing}},
    author = {A. Bagherzandi and S. Jarecki and Y. Lu and N. Saxena},
    booktitle = {18th ACM Conference on Computer and Communications Security (CCS)},
    year = {2011},
    pages = {433-444}}

@inproceedings{oprf-ot-si,
    title = {{Efficient oblivious pseudorandom function with applications to adaptive OT and secure computation of set intersection}},
    author = {S. Jarecki and X. Liu},
    booktitle = {6th Theory of Cryptography Conference (TCC)},
    pages = {577-594},
    year = {2009}}

@inproceedings{pesto,
    author = {M. Freedman and Y. Ishai and B. Pinkas and O. Reingold},
    title = {{PESTO: Proactively secure distributed single sign-on, or how to trust a hacked server}},
    booktitle = {5th IEEE European Symposium on Security and Privacy (EuroS\&P)},
    year = {2020}}

@article{oprf-proved,
    author = {M. Naor and O. Reingold},
    title = {{Number-theoretic constructions of efficient pseudo-random functions}},
    journal = {Journal of the ACM},
  volume    = {51},
  number    = {2},
  pages     = {231-262},
  year      = {2004}}

@inproceedings{strong-oprf,
    author = {M. Freedman and Y. Ishai and B. Pinkas and O. Reingold},
    title = {{Keyword search and oblivious pseudorandom functions}},
    booktitle = {2nd Theory of Cryptography Conference (TCC)},
    year = {2005}}

@inproceedings{voprf-proved,
    author = {S. Jarecki and A. Kiayias and H. Krawczyk},
    title = {{Round-optimal password-protected secret sharing and T-PAKE in the password-only model}},
    booktitle = {AsiaCrypt},
    year = {2014}}

@inproceedings{oprf-bitcoin-wallet,
    author = {S. Jarecki and A. Kiayias and H. Krawczyk and J. Xu},
    title = {{Highly-efficient and composable password-protected secret sharing (or: How to protect your Bitcoin wallet online)}},
    booktitle = {1st IEEE European Symposium on Security and Privacy (EuroS\&P)},
    pages = {276-291},
    year = {2016}}

@misc{ZKP-BINF,
    author = {B. Diamond and J. Posen},
    title = {Succinct Arguments over Towers of Binary Fields},
    howpublished = {\url{https://eprint.iacr.org/2023/1784}},
    year = {2024}}

@inproceedings{blind-sign,
    author = {D. Chaum},
    title = {Blind Signatures for Untraceable Payments},
    booktitle = {CRYPTO},
    pages = {199-203},
    year = {1982}}

@article{privacypass,
    author = {A. Davidson and I. Goldberg and N. Sullivan and G. Tankersley and F. Valsorda},
    title = {{PrivacyPass: Bypassing Internet challenges anonymously}},
  journal   = {Privacy Enhancing Technologies},
  volume    = {2018},
  number    = {3},
  pages     = {164-180},
  year      = {2018}}

@inproceedings{anon-credential-2001,
    title={An efficient system for non-transferable anonymous credentials with optional anonymity revocation},
    author={Camenisch, J. and Lysyanskaya, A.},
    booktitle={EUROCRYPT},
    year={2001}}

@inproceedings{Opaak,
    author = {G. Maganis and E. Shi and H. Chen and D. Song},
    title = {{Opaak: Using mobile phones to limit anonymous identities online}},
    BOOKTITLE = {10th International Conference on Mobile Systems, Applications, and Services (MobiSys)},
    YEAR = {2012}}

@manual{uprov,
  title={{U-Prove technology overview v1.1}},
  author={C. Paquin},
  organization={Microsoft Corporation},
  year={2013}
}

@Misc{hyperledge-idemix,
    TITLE = {{MSP implementation with Identity Mixer}},
    author = {{Hyperledger Fabric}},
    note = {Accessed July 20, 2022},
howpublished = {\url{https://hyperledger-fabric.readthedocs.io/en/release-2.2/idemix.html}}}

@misc{trusttoken,
    title = {{TrustToken API}},
    author = {{Web Incubator CG}},
    note = {Accessed July 20, 2022},
    howpublished = {\url{https://github.com/WICG/trust-token-api}}}

@inproceedings{idemix,
    title = {{Design and implementation of the Idemix anonymous credential system}},
    author = {J. Camenisch and E. Herreweghen},
    booktitle = {9th ACM Conference on Computer and Communications Security (CCS)},
    year = {2002}}

@inproceedings{prima,
    author = {M. Asghar and M. Backes and M. Simeonovski},
    title = {{PRIMA: Privacy-preserving identity and access management at Internet-scale}},
    booktitle = {52nd IEEE International Conference on Communications (ICC)},
    year = {2018}}

@inproceedings{PseudoID,
    author = {A. Dey and S. Weis},
    title = {{PseudoID: Enhancing privacy for federated login}},
    booktitle = {3rd Hot Topics in Privacy Enhancing Technologies (HotPETs)},
    year = {2010}}

@inproceedings{anon-credential,
    author = {M. Chase and S. Meiklejohn and G. Zaverucha},
    title = {{Algebraic MACs and keyed-verification anonymous credentials}},
    booktitle = {21st ACM Conference on Computer and Communications Security (CCS)},
    year = {2014}}

@article{maler2008venn,
  author    = {E. Maler and
               D. Reed},
  title     = {{The venn of identity: Options and issues in federated identity management}},
  journal   = {IEEE Security \& Privacy},
  volume    = {6},
  number    = {2},
  pages     = {16-23},
  year      = {2008},
}

@inproceedings{BrowserID,
  author    = {D. Fett and
               R. K{\"{u}}sters and
               G. Schmitz},
  title     = {{Analyzing the BrowserID SSO system with primary identity providers
               using an expressive model of the Web}},
  booktitle = {20th European Symposium on Research in Computer Security (ESORICS)},
  year      = {2015}}

@inproceedings{HanCSTW18,
  author    = {J. Han and
               L. Chen and
               S. Schneider and
               H. Treharne and
               S. Wesemeyer},
  title     = {Anonymous Single-Sign-On for $n$ Designated Services with Traceability},
  booktitle = {23rd European Symposium on Research in Computer
               Security (ESORICS)},
  year      = {2018}}

@inproceedings{SPRESSO,
  author    = {D. Fett and
               R. K{\"{u}}sters and
               G. Schmitz},
  title     = {{SPRESSO: A secure, privacy-respecting single sign-on system for the Web}},
  booktitle = {22nd ACM Conference on Computer and Communications Security (CCS)},
  pages     = {1358-1369},
  year      = {2015},
}

@manual{OpenIDConnect,
  title={{OpenID Connect core 1.0 incorporating errata set 1}},
  author={Sakimura, N. and Bradley, J. and Jones, M. and de Medeiros, B. and Mortimore, C.},
  organization={The OpenID Foundation},
  year={2014}
}

@manual{OpenID-dy-regi,
  title={{OpenID Connect dynamic client registration 1.0 incorporating errata set 2}},
  author={Sakimura, N. and Bradley, J. and Jones, M.},
  organization={The OpenID Foundation},
  year={2023}
}

@misc{uber,
    title={{OIDC Web SDK}},
    note={Accessed April 10, 2025},
    author={{Uber Developers}},
    howpublished={\url{https://developer.uber.com/docs/consumer-identity/oidc/web}}}

@misc{GoogleIdIntegrate,
    howpublished={\url{https://developers.google.com/identity/gsi/web/guides/integrate/}},
title = {{Google Identity: Integration considerations}},
note = {Accessed January 13, 2025},
author = {{Google for Developers}}}

@misc{sop,
    AUTHOR = {{W3C Web Security}},
    title = {Same Origin Policy},
    howpublished = {\url{https://www.w3.org/Security/wiki/Same_Origin_Policy}},
}

@misc{htmlStorageHtml,
    author = {{The WHATWG Community}},
    title = {{HTML} living standard: 12 Web storage},
    howpublished = {\url{https://html.spec.whatwg.org/multipage/webstorage.html}},
}

@misc{postm-targeto,
    author = {{The WHATWG Community}},
    title = {{HTML} living standard: 9.3 Cross-document messaging},
    howpublished = {\url{https://html.spec.whatwg.org/multipage/web-messaging.html}},
}

@inproceedings{FettKS14,
  author    = {D. Fett and
               R. K{\"{u}}sters and
               G. Schmitz},
  title     = {An Expressive Model for the Web Infrastructure: {Definition} and Application
               to the {BrowserID} {SSO} System},
  booktitle = {35th {IEEE} Symposium on Security and Privacy (S{\&}P)},
  year      = {2014}}

@manual{SAML,
  title={{Profiles for the OASIS security assertion markup language (SAML) v2.0}},
  author={Hughes, J. and Cantor, S. and Hodges, J. and Hirsch, F. and Mishra, P. and Philpott, R. and Maler, E.},
  organization={OASIS},
  year={2005}
}

@manual{SAMLIdentifier,
  title={{SAML V2.0 subject identifier attributes profile version 1.0}},
  author={T. Hardjono and S. Cantor},
  organization={OASIS},
  year={2018}
}

@manual{NIST2017draft,
  title={{SP 800-63C: Digital identity guidelines: Federation and assertions}},
  author={P. Grassi and E. Nadeau and J. Richer and S. Squire and J. Fenton and N. Lefkovitz and J. Danker and Y.-Y. Choong and K. Greene and M. Theofanos},
  organization={National Institute of Standards and Technology (NIST)},
  year={2017}
}

@Misc{MITREid,
author = {J. Richer},
howpublished = {\url{http://mitreid-connect.github.io/index.html}},
title = {{MITREid Connect v1.3.3}},
note = {Accessed August 20, 2021}
}

@Misc{elliptic-lib,
author = {Fedor Indutny},
howpublished = {\url{https://www.npmjs.com/package/elliptic}},
title = {{Fast elliptic curve cryptography in plain JavaScript}},
note = {Accessed July 20, 2025}
}

@Misc{FirefoxAccount,
howpublished = {\url{https://mozilla.github.io/application-services/docs/accounts/welcome.html}},
note = {Accessed August 20, 2019},
title = {{About Firefox Accounts}},
author = {Firefox Application Services}
}

@inproceedings{WangWS13,
  author    = {J. Wang and
               G. Wang and
               W. Susilo},
  title     = {Anonymous Single Sign-On Schemes Transformed from Group Signatures},
  booktitle = {5th International Conference on Intelligent Networking and Collaborative
               Systems (INCoS)},
  year      = {2013}}

@inproceedings{dimvaLiM16,
  author    = {W. Li and
               C. Mitchell},
  title     = {Analysing the security of {Google's} implementation of {OpenID} {Connect}},
  booktitle = {13th International Conference on Detection of Intrusions and Malware {\&} Vulnerability Assessment (DIMVA)},
  year      = {2016}}

@manual{de2014oauth,
  title={{OAuth 2.0 multiple response type encoding practices}},
  author={de Medeiros, B. and Scurtescu, M. and Tarjan, P. and Jones, M.},
  organization={The OpenID Foundation},
  year={2014}
}

@article{HanCSTWW20,
  author    = {J. Han and
               L. Chen and
               S. Schneider and
               H. Treharne and
               S. Wesemeyer and
               N. Wilson},
  title     = {Anonymous Single Sign-On With Proxy Re-Verification},
  journal   = {IEEE Transactions on Information Forensics and Security},
  volume    = {15},
  pages     = {223-236},
  year      = {2020}
}

@article{ELPASSO,
  author    = {Z. Zhang and
               M. Kr{\'{o}}l and
               A. Sonnino and
               L. Zhang and
               E. Rivi{\`{e}}re},
  title     = {{EL PASSO: Efficient and lightweight privacy-preserving single sign on}},
  journal   = {Privacy Enhancing Technologies},
  volume    = {2021},
  number    = {2},
  pages     = {70-87},
  year      = {2021},
}

@inproceedings{UnlimitID,
  author    = {M. Isaakidis and
               H. Halpin and
               G. Danezis},
  title     = {{UnlimitID: Privacy-preserving federated identity management using
               algebraic MACs}},
  booktitle = {15th ACM Workshop on Privacy in the Electronic Society (WPES)},
  pages     = {139-142},
  year      = {2016},
}

@manual{referer_policy,
author = {J. Eisinger and E. Stark},
  title = {{W3C candidate recommendation: Referrer policy}},
  organization = {World Wide Web Consortium (W3C)},
  year = 2017,
}

@manual{irtf-cfrg-hash-to-curve-16,
  organization={Internet Engineering Task Force},
        author =	{A. Faz-Hernandez and S. Scott and N. Sullivan and R. Wahby and C. Wood},
	title =		{draft-irtf-cfrg-hash-to-curve-16: Hashing to elliptic curves},
    year = {2022}
}

@manual{rfc6749,
  author    = {D. Hardt},
  title     = {{RFC 6749: The OAuth 2.0 authorization framework}},
  organization={Internet Engineering Task Force},
  year      = {2012},
}

@manual{ohttp-rfc,
  author    = {M. Thomson and C. Wood},
  title     = {{RFC 9458: Oblivious HTTP}},
  organization={Internet Engineering Task Force},
  year      = {2024},
}

@inproceedings{zkp-benchmark,
    title = {{zk-Bench: A toolset for comparative evaluation and performance benchmarking of SNARKs}},
    author = {J. Ernstberger and S. Chaliasos and G. Kadianakis and S. Steinhorst and P. Jovanovic and A. Gervais and B. Livshits and M. Orru},
    booktitle = {14th International Conference on Security and Cryptography for Networks (SCN)},
    YEAR = {2024}}

@inproceedings{ring-sig,
    title = {Ring signatures: Stronger definitions, and constructions without random oracles},
    author = {A. Bender and J. Katz and R. Morselli},
    pages = {60-79},
    year = {2006},
    booktitle = {3rd Theory of Cryptography Conference (TCC)}}

@inproceedings{tor,
    title = {Tor: The Second-Generation Onion Router},
    author = {R. Dingledine and N. Mathewson  and P. Syverson},
    pages = {303-320},
    year = {2004},
    booktitle = {13th USENIX Security Symposium}}




\end{document}